\documentclass[lettersize,journal]{IEEEtran}

\pdfoutput=1
\usepackage{amsmath,amsfonts}
\usepackage{algorithmic}
\usepackage{algorithm}
\usepackage{array}
\usepackage{textcomp}
\usepackage{stfloats}
\usepackage{url}
\usepackage{verbatim}
\usepackage{graphicx}
\usepackage{cite}
\usepackage[nolist]{acronym}
\usepackage{booktabs}
\usepackage{multirow}
\usepackage{enumitem}
\usepackage{epstopdf}
\usepackage{bm}
\usepackage{tabularx}
\usepackage{amssymb}
\usepackage{arydshln}
\usepackage{svg}
\usepackage{rotating}
\usepackage{tikz}
\usepackage{placeins }
\usepackage{array}
\usepackage{caption}
\usepackage{subcaption}
\usepackage{xcolor}

\usepackage[colorlinks, citecolor=blue]{hyperref}
\begin{document}
\newacro{HRI}[HRI]{Human-Robot Interaction}
\newacro{VSSE}[VSSE]{Visual Scene-aware Speech Enhancement}
\newacro{VSE}[VC-S$^2$E]{Visual-prompting ConMamba for Scene-aware Speech Enhancement}
\newacro{SNR}[SNR]{Signal-to-Noise Ratio}
\newacro{SAVSE}[SAV-SE]{Scene-aware Audio-Visual Speech Enhancement}
\newacro{NMF}[NMF]{Non-negative Matrix Factorization}
\newacro{DNN}[DNN]{Deep Neural Network}
\newacro{CNN}[CNN]{Convolutional Neural Network}
\newacro{GAN}[GAN]{Generative Adversarial Network}
\newacro{RNN}[RNN]{Recurrent Neural Network}
\newacro{DCCRN}[DCCRN]{Deep Complex Convolution Recurrent Network}
\newacro{CAV-MAE}[CAV-MAE]{Contrastive Audio-Visual Masked Autoencoder}
\newacro{STFT}[STFT]{Short-Time Fourier Transform}
\newacro{ISTFT}[ISTFT]{Inverse Short-Time Fourier Transform }
\newacro{SSM}[SSM]{State Space Model}
\newacro{BiMamba}[BiMamba]{Bidirectional Mamba}
\newacro{SA}[SA]{Self-attention}
\newacro{GCAM}[Grad-CAM]{Gradient-weighted Class Activation Mapping}
\newacro{ConMamba}[ConMamba]{Conformer Bimamba}
\newacro{HMMs}[HMMs]{Hidden Markov Models}
\newacro{PSM}[PSM]{Phase-Sensitive Mask}
\newacro{SE}[SE]{Speech Enhancement}
\newacro{AVSE}[AVSE]{Audio-Visual Speech Enhancement}
\newacro{PESQ}[PESQ]{Perceptual Evaluation of Speech Quality}
\newacro{STOI}[STOI]{Short-Time Objective Intelligibility}
\newacro{MSE}[MSE]{Mean-Square Error}
\newacro{BMamba}[BiMamba]{Bidirectional Mamba}
\newacro{DL}[DL]{Deep Learning}
\newacro{ASR}[ASR]{Automatic Speech Recognition}
\newacro{GLU}[GLU]{Gated Linear Unit}
\newacro{TCN}[TCN]{Temporal Convolutional Network}
\newacro{MHSA}[MHSA]{multi-Head Self-Attention}
\newcommand{\concat}{\textcircled{c}}

\title{SAV-SE: Scene-aware Audio-Visual Speech Enhancement with Selective State Space Model}

\author{Xinyuan Qian,~\IEEEmembership{Senior Member,~IEEE,}
Jiaran Gao, 
Yaodan Zhang,
Qiquan Zhang,~\IEEEmembership{Member,~IEEE,}
Hexin Liu, 
Leibny Paola Garcia,~\IEEEmembership{Senior Member,~IEEE,} 
Haizhou Li,~\IEEEmembership{Fellow,~IEEE,}
\thanks{
Manuscript received date; revised date. 
This work was supported in part by National Natural Science Foundation of China under Grant No. 62306029, Beijing Natural Science Foundation under Grants L233032, in part by Shenzhen Research Institute of Big Data under Grant No. K0012024000, in part by Shenzhen Science and Technology Program (Shenzhen Key Laboratory Grant No. ZDSYS20230626091302006), and in part by the Program for Guangdong Introducing Innovative and Enterpreneurial Teams, Grant No. 2023ZT10X044. \textit{(Corresponding Author: Qiquan Zhang)}.
}
\thanks{Xinyuan Qian, Jiaran Gao, Yaodan Zhang are with the School of Computer and Communication Engineering, University of Science and Technology Beijing, Beijing, 100083, China (e-mail: qianxy@ustb.edu.cn, u202241904@xs.ustb.edu.cn, 
u202241925@xs.ustb.edu.cn).}
\thanks{Qiquan Zhang is with the School of Electrical Engineering and Telecommunications, The University of New South Wales, Sydney, 2052, Australia~(e-mail: {zhang.qiquan}@outlook.com).}
\thanks{Hexin Liu is with the College of Computing and Data Science, Nanyang Technological University, Singapore (email: hexin.liu@ntu.edu.sg).}
\thanks{Leibny Paola Garcia is with the Center for Language and Speech Processing, Johns Hopkins University, USA (e-mail: lgarci27@jhu.edu).}
\thanks{Haizhou Li is with the Guangdong Provincial Key Laboratory of Big Data Computing, The Chinese University of Hong Kong (Shenzhen), 518172 China, and also with Shenzhen Research Institute of Big data, Shenzhen, 51872 China (e-mail: {haizhouli}@cuhk.edu.cn). 
}
}

\markboth{Journal of \LaTeX\ Class Files,~Vol.~14, No.~8, August~2021}%
{Shell \MakeLowercase{\textit{et al.}}: A Sample Article Using IEEEtran.cls for IEEE Journals}


\maketitle

\begin{abstract}
Speech enhancement plays an essential role in various applications, and the integration of visual information has been demonstrated to bring substantial advantages. However, the majority of current research concentrates on the examination of facial and lip movements, which can be compromised or entirely inaccessible in scenarios where occlusions occur or when the camera view is distant. Whereas contextual visual cues from the surrounding environment have been overlooked: for example, when we see a dog bark, our brain has the innate ability to discern and filter out the barking noise. To this end, in this paper, we introduce a novel task, i.e. \ac{SAVSE}. To our best knowledge, this is the first proposal to use rich contextual information from synchronized video as auxiliary cues to indicate the type of noise, which eventually improves the speech enhancement performance.  Specifically, we propose the VC-S$^2$E method, which incorporates the Conformer and Mamba modules for their complementary strengths. Extensive experiments are conducted on public MUSIC, AVSpeech and AudioSet datasets, where the results demonstrate the superiority of VC-S$^2$E over other competitive methods. We will make the source code publicly available. Project demo page: https://AVSEPage.github.io/
\end{abstract}


\section{Introduction}\label{sec:intro}
\IEEEPARstart{I}{n} {our daily living environments, speech signals are often distorted by various environmental background noises during their propagation. Speech enhancement (SE) is a task aiming at isolating the clean speech in the presence of noise interference, resulting in improved speech intelligibility and perceptual quality~\cite{loizou,overview2018,borgstrom2010improved,xian2020multi}.} It enables natural and effective \ac{HRI} and plays a crucial role in various applications, such as hearing aids, mobile communication, automatic speech recognition~\cite{gulati2020conformer,wang2022predict,rodomagoulakis2019improved,liu2024aligning}, speaker verification~\cite{wan2018generalized}, and speaker tracking~\cite{qian2022audio,qian2021audio,qian2021multi}. These applications underscore the importance of SE in realistic scenarios. Traditional signal processing-based SE approaches, which are derived from the assumed properties on speech and noise, are incapable of suppressing highly non-stationary noise sources~\cite{mmse,2007mmse,zhang2019,LogNC}. In the past decade, with the advent of deep learning technology and increased computational resources, supervised speech enhancement solutions has achieved great success~\cite{overview2018}.

\begin{figure}[!t] 
\centering
\includegraphics[width=\columnwidth]{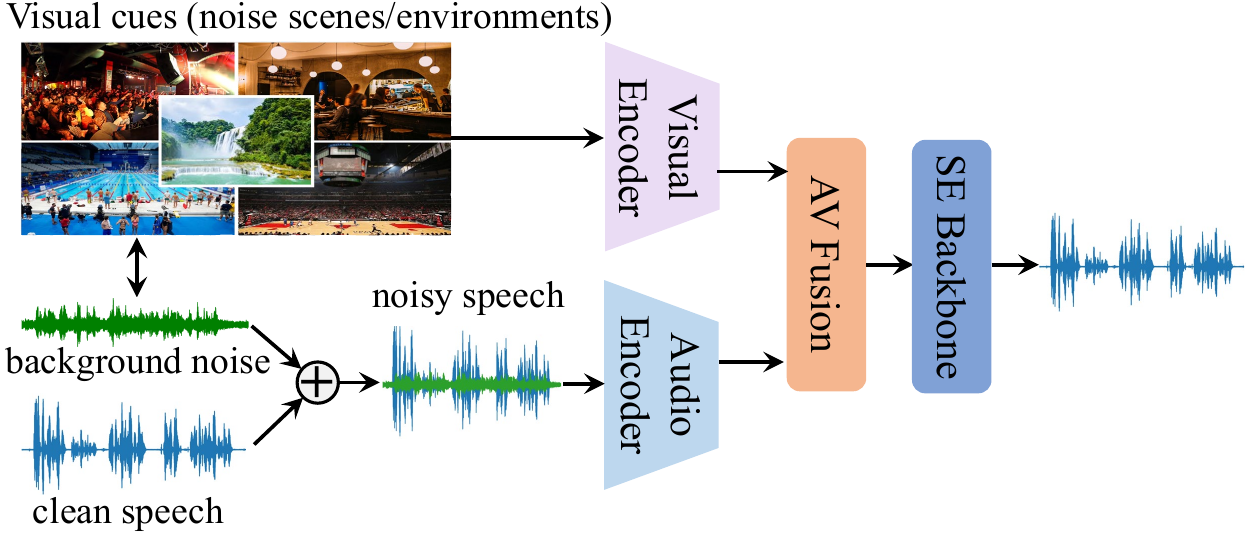}
\caption{Our proposed \ac{SAVSE} task where outputs from the audio and visual encoder are fused to refine and generate the enhanced audio. 
By incorporating the visual context from noise environments, it significantly enhances speech quality, particularly in situations where traditional audio-only techniques falter.}
\label{fig:intro}
\end{figure}


Despite the significant strides made in the field, the challenge of noise reduction without inflicting artifacts on the speech signal persists, particularly in dynamic environments characterized by non-stationary and multi-source noise~\cite{sivaraman2022efficient}. This difficulty is further compounded by the need to maintain the integrity of the speech signal, ensuring that the naturalness of the human voice is preserved.

To address this challenge, researchers have been exploring cutting-edge signal processing methodologies and sophisticated machine learning paradigms. One promising solution involves the use of neural networks, which has demonstrated great capabilities in extracting features and separating signals from complex acoustic environments. 
A variety of network architectures
are trained to learn the underlying patterns in noisy audio data, thus accomplishing the objective of speech enhancement~\cite{taha2018survey}.
Each of these models contributes unique strengths to the task of learning and generalizing from noisy audio data.
For example, Multi-Layer Perceptrons (MLPs) are proficient in detecting intricate, non-linear data patterns, whereas \ac{RNN} effectively manage the sequential dependencies in audio signals. \ac{TCN} excel in capturing long-range dependencies without suffering from the vanishing gradient problem that plagues standard \ac{RNN}. The Transformer architecture, featuring self-attention, has transformed the field by allowing models to process any part of the input sequence, which is crucial for tasks involving widespread noise-speech relationships. The Mamba architecture~\cite{mambaspeech}, as the latest advancement, further extends the capabilities of noise reduction and speech enhancement.

Researchers have increasingly acknowledged the importance of maintaining semantic, temporal, and spatial coherence between audio and video sources~\cite{wei2022learning,lin2021exploiting}.
This motivates attempts to use video information as a complement of audio input to recover details that are lost in audio-only scenarios. Existing \ac{AVSE} schemes often exploit temporal synchronized facial and lip movements to improve the clarity and perception of enhanced speech~\cite{gabbay2017visual,afouras2018conversation,gabbay2018seeing}. 
Despite outperforming audio-only SE systems, they are infeasible in many practical scenarios (e.g., outdoors or pandemic period) where human visual cues are not available. 
Moreover, inaccurate face or lip detection (e.g., in low-quality videos) may also result in degraded performance. 
In contrast, visual cues of environmental information, such as noise scenes or background objects emitting the noise, are easier to capture. It is more practical to use visual environmental cues to provide a valuable complement to speech enhancement.
Thus, to fully leverage audio-visual information to enhance uni-modal learning, it is essential to consider these modality-specific attributes.

\IEEEpubidadjcol 
In this paper, we introduce a novel \ac{AVSE} framework, as illustrated in Figure~\ref{fig:intro}, which uses visual information of the surrounding scenes as an auxiliary prompt to improve SE performance. Specifically, it addresses the limitations of current technologies, particularly in scenarios where an accurate capture of facial or lip information is not available.

The contributions of this paper are summarized as follows:
\begin{enumerate}[label=\arabic*.]
    \item We introduce a novel and more practical scene-aware AVSE task, namely \ac{SAVSE}. Unlike existing \ac{AVSE} studies that rely primarily on visual facial and lip movements, this paper explores auxiliary visual contextual cues from the surrounding scenes to mitigate environmental background noise.
    \item We are the first to explore selective State Space Model (SSM) for audio-visual speech enhancement. 
    Specifically, we propose a \ac{VSE}, a novel approach that leverages audio-visual modalities to improve speech quality and intelligibility. Built upon innovative hybrid convolution-SSM architecture, ConMamba can capture both long-range global interactions and localized fine-grained feature patterns. 
    \item We comprehensively evaluate our proposed method across three widely used AV datasets. The results consistently confirm the superiority of our $\text{VC-}\text{S}^{2}\text{E}$ over other competing methods in speech quality and intelligibility. Meanwhile, the visualization analysis illustrates that visual focal areas locate at the sounding object, demonstrating the contribution of visual scene information.
    
\end{enumerate}

\section{Related Work}\label{s1ec:relatedwork}
This section describes the evolution of speech enhancement, from traditional to state-of-the-art audio-visual and state space model techniques,  assessing their impact on speech processing and their capacity to overcome current limitations.

\subsection{Audio only Speech Enhancement}

Traditional SE mainly include Wiener filtering~\cite{wiener1996}, spectral subtraction~\cite{PALIWAL2010450}, statistical model methods~\cite{mmse2017,zhang2019}, and \ac{NMF} algorithms~\cite{IMM2006-04511}. Due to assumptions about the statistical characteristics of speech and noise, these methds fail to handle highly non-stationary noise signals. In the past decades, \ac{DL} has revolutionized the field of SE, demonstrating impressive performance over traditional schemes~\cite{overview2018,purwins2019deep}.

\ac{DL}-based SE methods can be broadly grouped into temporal domain and time-frequency (T-F) domain methods. Temporal domain methods, which perform denoising directly on the raw waveform, optimize a neural network to extract the clean waveform from noisy input~\cite{cleanunet,demcus,kolbaek2020loss}. In contrast, T-F domain methods separate clean speech from T-F representation, such as the magnitude spectrum~\cite{tfaj}, log-power spectrum~\cite{yongxu2015}, and complex spectrum~\cite{8910352}. 
Typically, a neural network is trained to map the noisy T-F representation to the clean T-F representation or a multiplicative T-F mask. 

Different network architectures have been investigated for SE. With the ability to capture long-range dynamic dependencies, the Long-Short Term Memory (LSTM) network has shown substantial performance gains over the MLP, especially in generalizing to unseen speakers~\cite{chenlstm}. Later, the \ac{TCN}, which uses stacked 1-D dilated convolution with residual connection, has demonstrated comparable or better performance than LSTM, with fewer parameters and faster training speed~\cite{GRN,DeepMMSE}. 
The Transformer learns global interactions effectively, enabling recent advances in SE~\cite{zhang2024empirical,cleanunet}. By combining convolutions with self-attention, Conformer~\cite{gulati2020conformer} learns global interactions while capturing fine-grained local characteristics to achieve state-of-the-art (SoTA) results~\cite{conformer-se}. 

{In addition to these advanced architectures, recent methods leverage explicit noise representation and adversarial training to tackle non-stationary or unseen noise. For instance, DNE\cite{Lee2020DynamicNE} employs VAD to embed noise features, NIT-CycleGAN \cite{ting2022speechenhancementbasedcyclegan} injects noise-type labels, and a two-stage VAE-GAN \cite{xiang2023twostagedeeprepresentationlearningbased} disentangles noise latent variables before adversarial refinement. Such designs significantly improve SE robustness in real-world noisy conditions.}

Most recently, the Mamba~\cite{gu2023mamba} neural architecture, a novel structured SSM has demonstrated great potential as an alternative to Transformer architecture in a variety of speech processing domains, including speech enhancement~\cite{mambaspeech,chao2024investigation} and speech separation~\cite{jiang2024dual}. There are further researches which investigate the capabilities of Mamba and other SSMs in SE, especially in scenarios where multi-modal data exist or of varying quality.

\subsection{{Audio-Visual Speech Enhancement}}


 Humans rely on multi-modal cues—such as visual and auditory signals—to explore, capture, and perceive the real world. These sensory modalities work together to provide a rich and integrated understanding of the environment. Over the past few decades, research has focused on investigating the semantic, temporal, and spatial consistencies between auditory and visual signals. Studies have shown that the complementary and mutually enhancing nature of these modalities can significantly improve perception and understanding in various contexts~\cite{wang2022self,tan2020audio}.
 For example, synchronizing audio and visual information has been shown to enhance object recognition, improve speech comprehension in noisy environments, and help more robust scene understanding in dynamic settings.
 In particular, video can be used as a supplement to discover details that are lost in acoustic scenarios. The resulting \ac{AVSE} methods offer a more robust and precise enhancement capability, exceeding what could be achieved with sound alone~\cite{michelsanti2021overview}.



The AVSS method \cite{gabbay2017visual} employs a dual-tower CNN architecture, where video frames (focusing on the speaker's mouth region) and audio spectrograms are encoded separately. A multi-modal mask is then generated to separate the target voice from the mixed audio, significantly improving speech enhancement performance by exploiting the spatial alignment between the visual and auditory signals.
VisualVoice \cite{gao2021visualvoice} uses the appearance of the speaker’s face as an additional visual prior to isolating the corresponding vocal qualities. By incorporating cross-modal matching and speaker consistency losses from unlabeled video data, this approach enhances speech enhancement (SE) performance, especially in scenarios where the audio signal is heavily degraded. The model benefits from visual information to maintain speaker consistency and refine the target speech signal. FlowAVSE \cite{jung2024flowavse} takes a novel approach by adopting a conditional flow matching algorithm and optimizing a diffusion-based U-net architecture~\cite{zhangemnlp}. This method enables high-quality speech generation based on cropped face images, demonstrating improved performance in real-world SE tasks. The integration of cooperative dual-attention mechanisms and dynamic audio-visual fusion strategies further enhances the robustness and effectiveness of AVSE, making it more adaptable to a variety of noisy environments. In~\cite{richter2023audio}, a score-based generative model is employed which leverages audio-visual embeddings from the self-supervised AV-HuBERT model to enhance speech quality and reduce artifacts like phonetic confusions. Specifically, it integrates layer-wise features from AV-HuBERT into a noise conditional score network, demonstrating improved performance and noise robustness. LA-VOCE~\cite{mira2023voce} employs a transformer-based architecture to predict mel-spectrograms from noisy audio-visual inputs, while the second stage uses a neural vocoder (HiFi-GAN) to convert these spectrograms into clear waveform audio under low SNR conditions. All the aforementioned methods rely on the effective usage of human facial information. However, the effective integration of facial information requires not only accurate facial feature extraction but also robust alignment between the visual and audio modalities under varying conditions.

\subsection{State Space Models}
State space models (SSMs) have emerged as a powerful alternative to Transformer-based architectures for modeling long-range contextual dependencies in sequential data. Unlike Transformers, which rely heavily on self-attention mechanisms to capture interactions between distant tokens, SSMs provide a more efficient framework for processing sequences by leveraging dynamic state transitions that model temporal or spatial dependencies. This approach enables them to handle longer-range dependencies with fewer computational resources and greater scalability. A recent advancement in this area is the introduction of Mamba \cite{gu2023mamba}, a selective state space model (SSM) designed to overcome some of the limitations of traditional SSMs.
Unlike Transformers, which rely on computationally expensive self-attention mechanisms scaling quadratically with sequence length, Mamba achieves linear scalability by leveraging efficient state-space transitions. This allows it to model long-range dependencies more effectively. Additionally, the selective scan mechanism of Mamba dynamically adjusts its state transition parameters based on input, enabling it to focus on task-relevant information while filtering out noise—a feature absent in TCNs and Transformers. Compared to TCNs, which are limited to capturing local temporal patterns, Mamba integrates global and localized features through its hybrid convolution-SSM design, balancing fine-grained detail and holistic sequence understanding.

The success of Mamba has extended beyond natural language processing (NLP) to other domains such as computer vision and multimodal processing. For example, the Vim model \cite{zhuvision} leverages bidirectional Mamba to capture dynamic global context, resulting in significant improvements in various vision tasks. Similarly, VMamba \cite{liu2024vmambavisualstatespace} introduces a cross-scan module, which enhances global context modeling by applying a four-way selective scan mechanism. In speech domain, the most recent study \cite{mambaspeech} explores Mamba as a   alternative to Transformer architectures in both causal and non-causal configurations, demonstrating its great potential to be the next-generation backbone for  ASR and SE. Specifically, Mamba's ability to model long-range dependencies and capture fine-grained temporal relationships makes it a compelling choice for these speech-related tasks, where context over long time horizons is crucial for accurate processing.

\vspace{-3mm}
\subsection{Summary}
Although previous AVSE methods have achieved remarkable success, they often encounter challenges such as visual input quality dependency, narrow focus on facial features, and limited use of environmental signals. 
To address these problems, we propose a novel \ac{VSE} method which is built on the selective SSM architecture.

Specifically, it incorporates visual environmental information as an auxiliary cue to conventional SE, which eventually improves speech perceptual quality and intelligibility in various scenarios.

\begin{figure*}[!tb]
    \centering
    \makebox[\textwidth]{\includegraphics[width=\textwidth]{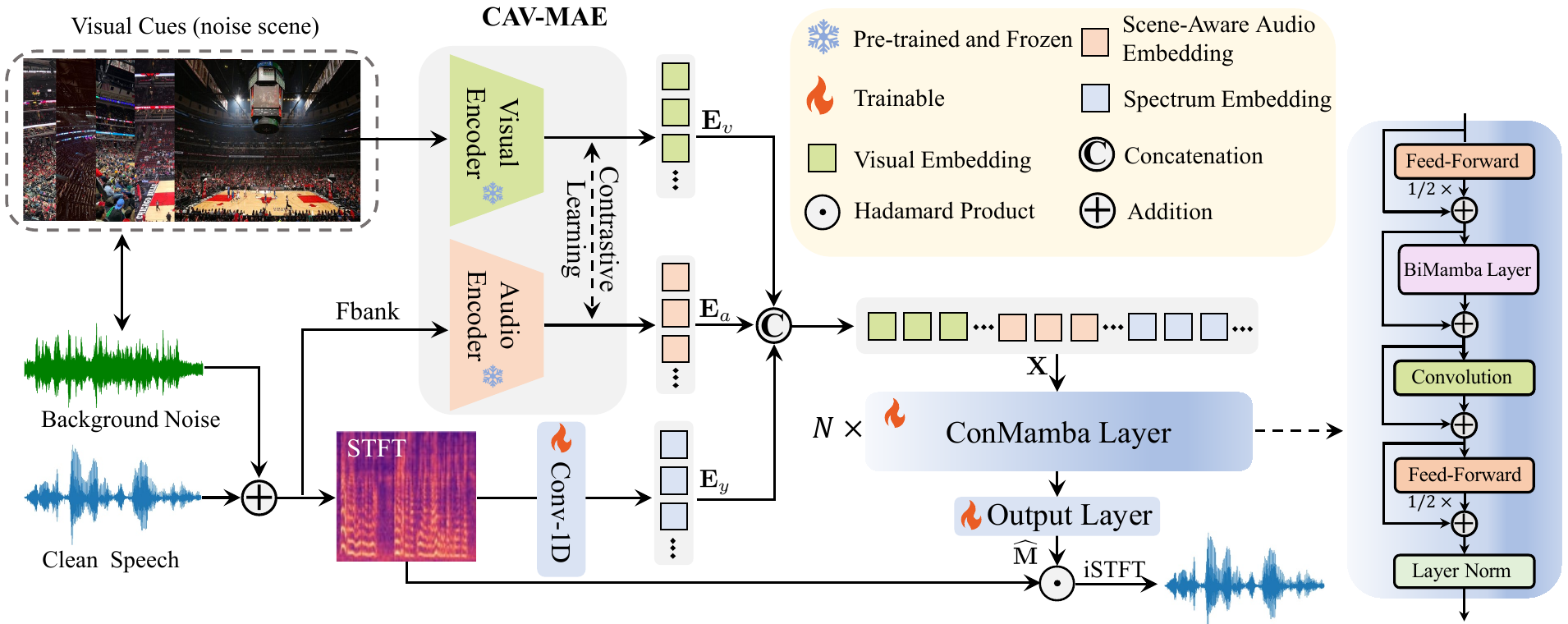}}  
    \caption{The block diagram of our proposed \ac{VSE} network which enhances human speech given the auxiliary visual cues from the noise scenario. }
    \label{fig:model2}
\end{figure*}

\section{Proposed \ac{VSE}}\label{sec:proposal}

\subsection{Problem Formulation}
\textcolor{black}{The observed noisy speech signals $\mathbf{y}\in\mathbb{R}^{1\times L}$ can be modeled as the addition of the clean speech $\mathbf{s}\in\mathbb{R}^{1\times L}$ and the noise $\mathbf{d}\in\mathbb{R}^{1\times L}$: $\mathbf{y}(l) = \mathbf{s}(l) + \mathbf{d}(l)$, $l={1,...,L}$. Let us denote $\mathbf{Y}_{t,k}$, $\mathbf{S}_{t,k}$, and $\mathbf{D}_{t,k}$ as the  short-time Fourier transform (STFT)  coefficients of $\mathbf{y}$, $\mathbf{s}$ and $\mathbf{d}$, where $t \in [1,T]$ and $k \in [1, K]$ index the time frames and frequency bins. A typical neural SE solution is to train a DNN to estimate a T-F mask $\mathbf{M}_{t,k}$. Here, we employ the phase-sensitive mask (PSM)~\cite{Erdogan2015PhasesensitiveAR}, defined as}

\begin{equation}
\mathbf{M}_{t,k} =
\frac{|\mathbf{S}_{t,k}|}{|\mathbf{Y}_{t,k}|}
\cos\left(\operatorname{arg}(\mathbf{S}_{t,k}) - \operatorname{arg}(\mathbf{Y}_{t,k})\right)
\end{equation}
where $|\cdot|$ extracts the magnitude, $\operatorname{arg}(\mathbf{S}_{t,k})$ and $\operatorname{arg}(\mathbf{Y}_{t,k})$ denote the spectral phase of the clean and noisy speech signals, respectively. 

Unlike existing AVSE methods using facial motions as the visual cues, which may be unavailable due to occlusions, off-camera targets, or low-resolution imagery, our approach leverages an image depicting the surrounding environment to provide noise prior for speech enhancement. This process is formulated as:
\begin{equation}
    \widehat{\mathbf{M}} =\mathbf{\Phi}(\mathbf{y},\mathbf{I}|\mathbf{\Omega})
\end{equation}
where $\mathbf{I}$ denotes the visual cue, $\mathbf{\Phi}$ denotes our proposed method SAV-SE network (see Sec.~\ref{sec:method}) with $\mathbf{\Omega}$ the trainable parameters.

The estimated mask is then applied to the noisy spectrum to attain the estimate of clean waveform $\widehat{\mathbf{s}}$ via inverse STFT (iSTFT):
\begin{equation}
\widehat{\mathbf{s}} = \text{iSTFT}(\widehat{\mathbf{M}} \odot \mathbf{Y})
\end{equation}
where $\odot$ denotes the element-wise multiplication.

\subsection{Method}\label{sec:method}
\textcolor{black}{In this section, we first describe the preliminaries of state space model (SSM). We then detail the workflow of our proposed \ac{VSE} and the network architecture.}

\subsubsection{Preliminaries} \textcolor{black}{The SSM originates from the Kalman filter, which takes a time-dependent set of inputs $u(t)\!\in\!\mathbb{R}$ and maps it into a set of outputs $y(t)\!\in\!\mathbb{R}$ through a hidden state $\mathbf{h}(t)\in\mathbb{R}^{N}$. Formally, the mapping process of SSMs can be represented as follows:
}
\begin{equation}\label{continue_equation}
\begin{aligned}
    \mathbf{h}'(t) &= \mathbf{A}\mathbf{h}(t) + \mathbf{B}u(t), \\
    y(t) &= \mathbf{C}\mathbf{h}(t),
\end{aligned}
\end{equation}
where $\mathbf{A}\!\in\!\mathbb{R}^{N\times N}$, $\mathbf{B}\!\in\!\mathbb{R}^{N\times 1}$, and $\mathbf{C}\!\in\!\mathbb{R}^{1\times N}$ are state matrix, input matrix, and output matrix, {representing continuous SSM parameters}. A discretization process is applied in advance to integrate SSM into deep learning architectures, where a timescale $\mathbf{\Delta}$ is used to transform matrices $\mathbf{A}, \mathbf{B}$ to their discrete counterparts $\overline{\mathbf{A}}, \overline{\mathbf{B}}$. A commonly employed technique for transformation is zero-order hold (ZOH), defined as:

\begin{equation}
\begin{aligned}
    \overline{\mathbf{A}} &= \exp{(\mathbf{\Delta}\mathbf{A})}, \\
    \overline{\mathbf{B}} &= (\mathbf{\Delta} \mathbf{A})^{-1} (\exp{\mathbf{\Delta} \mathbf{A}} - \mathbf{I})\cdot\mathbf{\Delta}\mathbf{B}.
    \end{aligned}
\end{equation}

Thus, Eq.~(\ref{continue_equation}) can be rewritten as:

\begin{equation}\label{mamba_euqation}
\begin{aligned}
    h_{t} & = \overline{\mathbf{A}}\mathbf{h}_{t-1} + \overline{\mathbf{B}}x_{t}, \\
    y_{t} &= \mathbf{C}\mathbf{h}_{t}.
\end{aligned}
\end{equation}

Furthermore, the output can be calculated via a global convolution:

\begin{equation}
\begin{aligned}
\overline{\mathbf{K}} & =\left(\mathbf{C} \overline{\mathbf{B}}, \mathbf{C} \overline{\mathbf{A B}}, \ldots, \mathbf{C} \overline{\mathbf{A}}^{L-1} \overline{\mathbf{B}}\right) \\
\mathbf{y} & =\mathbf{u} * \overline{\mathbf{K}},
\end{aligned}
\end{equation}
where $L$ is the length of the input sequence, $\overline{\mathbf{K}}$ is a structured convolution kernel, and $*$ denotes the convolution operation. Mamba introduces a selective scan mechanism, allowing the model to dynamically adjust $\mathbf{\Delta}$, $\mathbf{B}$, and $\mathbf{C}$ as functions of the input. This enables the model to learn dynamic representations while filtering out irrelevant information.

\subsubsection{Overview}
Figure~\ref{fig:model2} provides the \textcolor{black}{workflow} of our proposed \ac{VSE} which enhances human speech using the auxiliary visual cues from the noise scenario. Given the synchronized audio-visual streams, it first extracts the \textcolor{black}{visual scene} embedding \textcolor{black}{$\mathbf{E}_{v}\!\in\!\mathbb{R}^{T_v\times d_\text{model}}$}, the scenario-aware audio embedding \textcolor{black}{$\mathbf{E}_{a}\!\in\!\mathbb{R}^{T_a\times d_\text{model}}$} , and the acoustic spectrum embedding \textcolor{black}{$\mathbf{E}_{y}\!\in\!\mathbb{R}^{T_y\times d_\text{model}}$}. Then, it estimates the PSM mask to facilitate the reconstruction of the enhanced speech waveform $\widehat{\mathbf{s}}$ using the ConMamba backbone:
\begin{equation}
\widehat{\mathbf{M}} = \text{OLayer}(\text{ConMamba}(\mathbf{X};\Omega))
\end{equation}
where \textcolor{black}{$\mathbf{X} = [\mathbf{E}_{y}\ \concat\ \mathbf{E}_{a}\ \concat\ \mathbf{E}_{v}] \in \mathbb{R}^{(T_y+T_a+T_v)\times d_\text{model}}$} is the \textcolor{black}{fused} feature representation, $d_\text{model}$ represents the feature length, $T_v$, $T_a$ and $T_y$ indicate the {sequence lengths of the visual scene features, scenario-aware audio features, and acoustic spectrum features along the time dimension, respectively,}  and $\Omega$ is the trainable parameters; $\text{OLayer}(\cdot)$ indicates the output layer.

\subsubsection{Visual Encoder}  We employ the pre-trained contrastive audio-visual masked autoencoder (CAV-MAE)~\cite{gong2023contrastive}, which
integrates both contrastive learning and masked data modeling to jointly learn coordinated audio-visual representations.
 Specifically, the visual encoder is trained using contrastive loss, ensuring precise alignment between visual features (e.g., noisy scenes like vehicles and birds) and their corresponding audio features while distinguishing non-corresponding pairs.
To ensure temporal synchronization between video and audio, we use an re-sampling layer to match the video feature length to the audio one, which is crucial for their later effective fusion. The resulting visual features are formulated as:
\begin{equation}\label{eq:visualfeature_cavmae}
    \mathbf{E}_{v}= \text{CAV-MAE}^v({\mathbf{I}})
\end{equation}
\textcolor{black}{where $\text{CAV-MAE}^v(\cdot)$ and $\mathbf{I}$ denote the video encoder and the image input, respectively.}

\subsubsection{Audio Encoder}
 The pretrained CAV-MAE enhances the robustness of audio and visual features, making it particularly effective for multi-modal tasks that require a deep understanding of cross-modal interaction.
To capture the temporal and semantic consistencies between audio and video signals, 
\textcolor{black}{we adopt the pretrained audio encoder to extract the visual scenario-aware audio embedding:
\begin{equation}\label{eq:audiofeature_cavmae}
    \mathbf{E}_{a}= \text{CAV-MAE}^{a}(\mathbf{F})
\end{equation}
where $\text{CAV-MAE}^{a}(\cdot)$ and $\mathbf{F}$ denote the audio encoder and the Fbank acoustic feature of $\mathbf{y}$, respectively.}

The STFT magnitude spectrum $|{\mathbf{Y}}|$ is encoded using a 1D convolution layer to have same dimension with $\mathbf{E}_{v}$ and $\mathbf{E}_{a}$: 
\begin{equation}\label{eq:audiofeature_conv}
    \mathbf{E}_{y}= \text{Conv1D}(|{\mathbf{Y}}|)
\end{equation}

\subsubsection{ConMamba}
In Figure~2, the rectangle with gradient blue depicts the internal architecture of ConMamba in our proposed \ac{VSE}. Specifically, it consists of four stacked layers, which take the combined multi-modal feature representation ${\bf X}\in \mathbb{R}^{(T_y + T_a + T_v)\times d_\text{model}}$ as input to produce the PSM estimate ${\widehat{\bf M}}\in\mathbb{R}^{T\times K}$. 
{In conventional state space models, the hidden state is updated sequentially. In contrast, the BiMamba layer uses a data-dependent selective scanning mechanism, where core parameters are adaptively tuned based on the visual and audio content. When a visual token indicates a noise scene (e.g., an instrument or vehicle), bidirectional scanning updates the hidden states to associate the visual noise source with corresponding audio frequency features for targeted suppression. Additionally, depthwise separable 1D convolutions (DWConv) are integrated into each layer to precisely capture local interactions between adjacent visual and audio tokens.}

\textcolor{black}{Let us denote $\mathbf{Z}_{i}$ as the input to the $i$-th ConMamba layer.
It is initially processed by a feed-forward layer, followed by the BiMamba layer to capture bidirectional dependencies within the input sequence. Subsequently, a convolution layer is employed to learn fine-grained local feature pattern, leveraging pointwise convolutions and depthwise convolutions controlled by \ac{GLU}. Finally, another feed-forward layer is incorporated to solidify the learned transformations.}
This process can be formulated as:
\begin{equation}
\begin{aligned}
& \mathbf{Z}'_i = \mathbf{Z}_i + \frac{1}{2}\text{FFN}(\mathbf{Z}_{i}) \\
& \mathbf{Z}''_i = \mathbf{Z}'_i + \text{BiMamba}(\mathbf{Z}'_i) \\
& \mathbf{Z}'''_i = \mathbf{Z}''_i + \text{Conv}(\mathbf{Z}''_i) \\
& \mathbf{Z}^{out}_{i} = \text{LN}(\mathbf{Z}'''_i + \frac{1}{2}\text{FFN}(\mathbf{Z}'''_i))
\end{aligned}
\end{equation}
where $\mathbf{Z}^{out}_{i}$ is the output, FFN refers to the feed forward layer, BiMamba refers to the BiMamba layer, Conv refers to the convolution layer, and LN refers to layer normalization as described in the preceding sections.


\begin{figure}[!tb]
\centering
\includegraphics[width=\columnwidth]{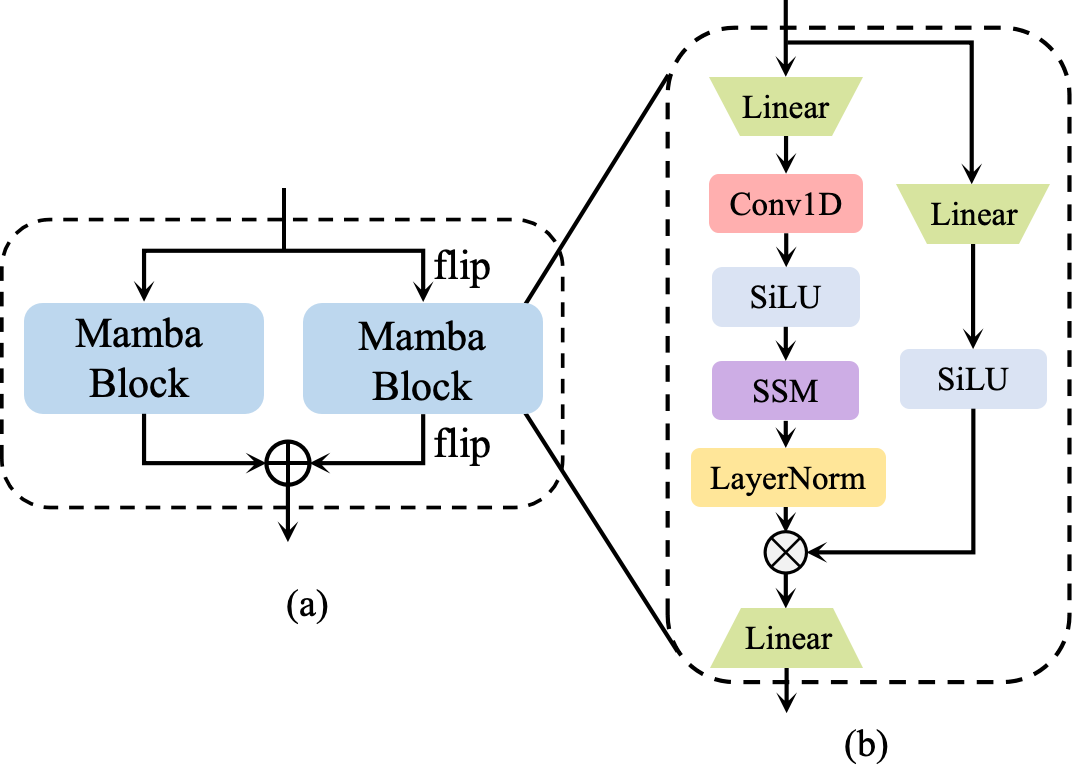}
\caption{{The illustrations of (a) the BiMamba layer and (b) the Mamba block.}}
\label{bimamba}
\vspace{-1.0em}
\end{figure}


{In particular, as illustrated in Figure~\ref{bimamba}, the BiMamba layer consists of two parallel Mamba blocks, each processing the input sequence along opposing temporal axes -- forward and backward, respectively.} Then, the bidirectional outputs are added, enabling the model to capture complex patterns and dependencies effectively. \textcolor{black}{Let $\mathbf{U}$ denote the input sequence to a BiMamba layer. Then, the forward and backward pass can be formulated as:}
\begin{equation}
\mathbf{h}_\text{f} = \text{Mamba}(\mathbf{U}) \quad \mathbf{h}_\text{b} = \text{Mamba}(\text{flip}({\mathbf{U}}))
\end{equation}
The bidirectional outputs are added to a layer normalization to provide output of the BiMamba layer:
\begin{equation}
\mathbf{Z}^{out} = \text{LN} \left( \mathbf{h}_\text{fused} + \mathbf{U} \right)
\end{equation}
where $\mathbf{h}_\text{fused} = \mathbf{h}_\text{f} + \text{flip}(\mathbf{h}_\text{b})$.

\textcolor{black}{\textbf{Mamba Block}. In Figure~\ref{bimamba} (b), we illustrate the detailed workflow of the Mamba block. Given the input feature representation denoted as $\mathbf{U}$, it is transformed through two parallel paths. The right path consists of a linear fully-connected layer followed by the SiLU activation function. The left path consists of a linear layer, a depthwise 1-D convolution unit, a SiLU activation, an SSM layer, and layer normalization in that sequence. The outputs of the two paths are then merged via element-wise multiplication, followed by a linear projection layer. Formally, the workflow of the Mamba block can be represented as:}
\begin{equation}
\begin{aligned}
& \mathbf{U}'_{1}  =\text{LN}\left(\text{SSM}\left(\text{SiLU}\left(\text{DWConv}\left(\text{Linear}\left(\mathbf{U}\right)\right)\right)\right)\right), \\
& \mathbf{U}'_{2} = \text{SiLU}\left(\text{Linear}\left(\mathbf{U}\right)\right),\\
& \mathbf{U}'' = \text{Linear}\left(\mathbf{U}'_{1}\odot \mathbf{U}'_{2} \right)
\end{aligned}
\end{equation}
where DWConv refers to the depthwise 1-D convolution. 

\section{Experiments}\label{sec:experiments}
\subsection{Experimental Setups}

\begin{table*}[!hbtp]
    \centering
    \small
    \setlength{\tabcolsep}{5.1pt} 
    \caption{Comparison results of our proposed \ac{VSE} method and the other approaches in PESQ ($\uparrow$) and STOI ($\uparrow$) with the SNR varies from -5 dB to 15 dB.}
    \begin{tabular}{@{}c l c ccccc c ccccc c@{}}
    \toprule[1.0pt]
    \multirow{2}{*}{\textbf{Dataset}} & \multirow{2}{*}{\textbf{Model}}  & \multicolumn{6}{c}{\textbf{PESQ}} & \multicolumn{6}{c}{\textbf{STOI} (in \%)} \\
    \cmidrule(lr){3-14}
    && \bf -5 dB & \bf 0 dB & \bf 5 dB & \bf 10 dB & \bf 15 dB & \bf Avg. & \bf -5 dB & \bf 0 dB & \bf 5 dB & \bf 10 dB & \bf 15 dB & \bf Avg. \\
    \midrule
    \multirow{9}{*}{\textbf{MUSIC}}
    & Unprocessed                   & 1.15 & 1.22 & 1.39 & 1.64 & 2.01 & 1.47 & 71.08 & 78.87 & 85.62 & 90.34 & 93.95 & 83.65 \\ 
    & MP-SENet                      & 1.77 & 2.15 & 2.59 & 2.90 & 3.29 & 2.54 & 85.02 & 89.61 & 92.88 & 94.68 & 96.07 & 91.65 \\ 
    & AViTAR                        & 1.95 & 2.36 & 2.78 & 3.09 & 3.46 & 2.73 & 86.96 & 90.94 & 93.94 & 95.51 & 96.70 & 92.81 \\
    & ExtBiMamba                    & 1.97 & 2.39 & 2.82 & 3.13 & 3.54 & 2.77 & 87.56 & 91.34 & 94.09 & 95.65 & 96.82 & 93.09 \\ 
    & AVDCNN                        & 1.45 & 1.70 & 2.12 & 2.88 & 3.00 & 2.23 & 83.62 & 84.72 & 90.86 & 95.05 & 94.88 & 89.83 \\
    & MLDM                          & 1.62 & 2.02 & 2.46 & 3.16 & 3.27 & 2.51 & 85.03 & 86.64 & 92.65 & 96.29 & 95.49 & 91.22 \\
    & ALTKD                         & 1.43 & 1.62 & 1.88 & 2.61 & 2.70 & 2.05 & 78.32 & 83.47 & 89.24 & 94.16 & 93.95 & 87.83 \\
    & DNE                           & 2.03 & 2.44 & 2.90 & \textbf{3.65} & 3.63 & 2.93 & 89.51 & 90.06 & 93.99 & \textbf{97.21} & 96.28 & 93.41 \\
    & \textbf{\ac{VSE}(our)}        & \textbf{2.23} & \textbf{2.66} & \textbf{3.08} & 3.38 & \textbf{3.73} & \textbf{3.02} & \textbf{89.55} & \textbf{92.61} & \textbf{94.96} & 96.24 & \textbf{97.18} & \textbf{94.11} \\
    \midrule
    \midrule
    \multirow{9}{*}{\textbf{AVSpeech}}
    & Unprocessed                   & 1.10 & 1.18 & 1.29 & 1.53 & 1.98 & 1.44 & 63.21 & 73.37 & 82.42 & 89.01 & 93.54 & 81.11 \\ 
    & MP-SENet                      & 1.44 & 1.76 & 2.06 & 2.57 & 2.97 & 2.16 & 80.74 & 86.25 & 88.97 & 91.36 & 93.42 & 88.15 \\ 
    & AViTAR                        & 1.63 & 1.99 & 2.30 & 2.82 & 3.22 & 2.39 & 83.16 & 88.05 & 90.36 & 92.96 & 94.71 & 89.85 \\
    & ExtBiMamba                    & 1.67 & 2.05 & 2.38 & 2.88 & 3.24 & 2.44 & 84.64 & 88.78 & 91.01 & 92.77 & 94.15 & 90.27 \\ 
    & AVDCNN                        & 1.18 & 1.28 & 1.69 & 2.22 & 2.57 & 1.79 & 68.29 & 73.47 & 86.47 & 92.04 & 92.50 & 82.55 \\
    & MLDM                          & 1.32 & 1.69 & 2.10 & 2.68 & 2.79 & 2.12 & 74.77 & 80.62 & 90.44 & 94.03 & 93.85 & 86.74 \\
    & ALTKD                         & 1.13 & 1.25 & 1.59 & 2.08 & 2.31 & 1.67 & 63.92 & 75.12 & 84.72 & 91.26 & 91.67 & 81.34 \\
    & DNE                           & 1.68 & 2.09 & 2.51 & 3.01 & 3.29 & 2.52 & 83.74 & 88.22 & 92.64 & \textbf{95.46} & 95.05 & 91.02 \\
    & \textbf{\ac{VSE}(our)}        & \textbf{1.96} & \textbf{2.39} & \textbf{2.72} & \textbf{3.20} & \textbf{3.53} & \textbf{2.76} & \textbf{88.13} & \textbf{90.86} & \textbf{92.90} & 95.24 & \textbf{96.38} & \textbf{92.70} \\
    \midrule
    \midrule
    \multirow{9}{*}{\textbf{AudioSet}} 
    & Unprocessed                   & 1.10 & 1.20 & 1.33 & 1.52 & 1.88 & 1.41 & 62.28 & 74.14 & 85.23 & 88.63 & 93.95 & 80.99 \\ 
    & MP-SENet                      & 1.33 & 1.75 & 2.19 & 2.49 & 2.92 & 2.14 & 75.85 & 86.76 & 92.33 & 93.94 & 96.07 & 88.99 \\ 
    & AViTAR                        & 1.42 & 1.89 & 2.43 & 2.71 & 3.14 & 2.32 & 78.30 & 87.89 & 93.46 & 94.65 & 97.14 & 90.29 \\
    & ExtBiMamba                    & 1.60 & 2.06 & 2.64 & 2.92 & 3.30 & 2.50 & 81.43 & 89.57 & 94.34 & 95.48 & 97.48 & 91.66 \\
    & AVDCNN                        & 1.25 & 1.43 & 1.65 & 2.17 & 2.70 & 1.84 & 70.24 & 82.52 & 85.35 & 88.95 & 94.56 & 84.32 \\
    & MLDM                          & 1.43 & 1.78 & 1.81 & 2.34 & 2.78 & 2.03 & 76.67 & 85.53 & 87.66 & 90.50 & 95.50 & 87.17 \\
    & ALTKD                         & 1.18 & 1.41 & 1.53 & 1.97 & 2.42 & 1.70 & 65.74 & 80.79 & 83.73 & 87.86 & 93.72 & 82.37 \\
    & DNE                           & 1.67 & \textbf{2.12} & 2.26 & 2.84 & 3.25 & 2.43 & 81.02 & 88.95 & 90.99 & 92.22 & 96.39 & 89.91 \\    
    & \textbf{\ac{VSE}(our)}        & \textbf{1.73} & 2.10 & \textbf{2.73} & \textbf{3.00} & \textbf{3.39} & \textbf{2.59} & \textbf{83.08} & \textbf{90.03} & \textbf{94.78} & \textbf{95.77} & \textbf{97.61} & \textbf{92.25} \\
    \toprule[1.0pt]
    \end{tabular}
    \label{tab:comparison_results}
\end{table*}

\subsubsection{Datasets} 
For clean speech data, we employ the  LibriSpeech \textit{train-clean-100} subset~\cite{7178964}, containing $28\,539$ utterances from $251$ speakers. For noise data including noisy recording and the corresponding visual scene, we employ three commonly used datasets: 
MUSIC~\cite{zhao2019sound}, AVSpeech~\cite{ephrat2018looking}, and AudioSet~\cite{audioset}.
Specifically, 
{the MUSIC dataset captures structured noise from musical instruments in controlled environments, AVSpeech provides diverse multi-speaker and multi-language scenarios with varied face poses, and AudioSet covers extensive environmental sounds, including both natural and human-made noises. 
These characteristics collectively align with the diverse conditions where SAV-SE is designed to excel.
These three datasets cover various aspects of audio environments and provide a comprehensive evaluation.
The data used in training and testing are non-overlapping, with each recording featuring unseen paired visual scenes and noisy sequences.}

\subsubsection{Baselines}

We choose seven different state-of-the-art methods to compare with our proposed \ac{VSE} algorithm, including
i) Conformer-based MP-SENet~\cite{mpsenet} adopts a codec architecture for SE where the encoder and decoder are bridged by Conformer.
ii) AViTAR~\cite{chen22vam} is an image-depicted sound generation method that transforms sounds from one space to another by altering their scene-driven acoustic signatures. We re-implement their method to fit our proposed \ac{SAVSE} task. 
iii) ExtBiMamba~\cite{mambaspeech} is the first exploration of \ac{BMamba} in  monaural SE. Specifically, it demonstrates the superiority of BiMamba as an alternative to Transformer and Conformer in SE.
{
iv) AVDCNN \cite{hou2018audio}
uses a dual-stream CNN with late fusion for multi-task audio-visual learning, enhancing speech via visual cues.
v) MLDM \cite{10508446} introduces a unified multi-level distortion measure for comprehensive speech enhancement.
vi) ALTKD \cite{Zheng_2023}
uses a U-Net-based teacher-student model, where the teacher fuses ultrasound tongue images, lip videos, and audio, distilling tongue information for high-quality enhancement with only audio and lip video needed at inference.
vii) DNE \cite{Lee2020DynamicNE}
leverages VAD to capture noise characteristics for improved SE 
in non-stationary and unseen noise conditions.
}
It should be noted that, for fair comparison, all baselines are evaluated under the same parameter settings.

\begin{figure*}[!tb]
\centering
\begin{subfigure}[t]{.49\linewidth}
    \centering
    \includegraphics[width=\linewidth]{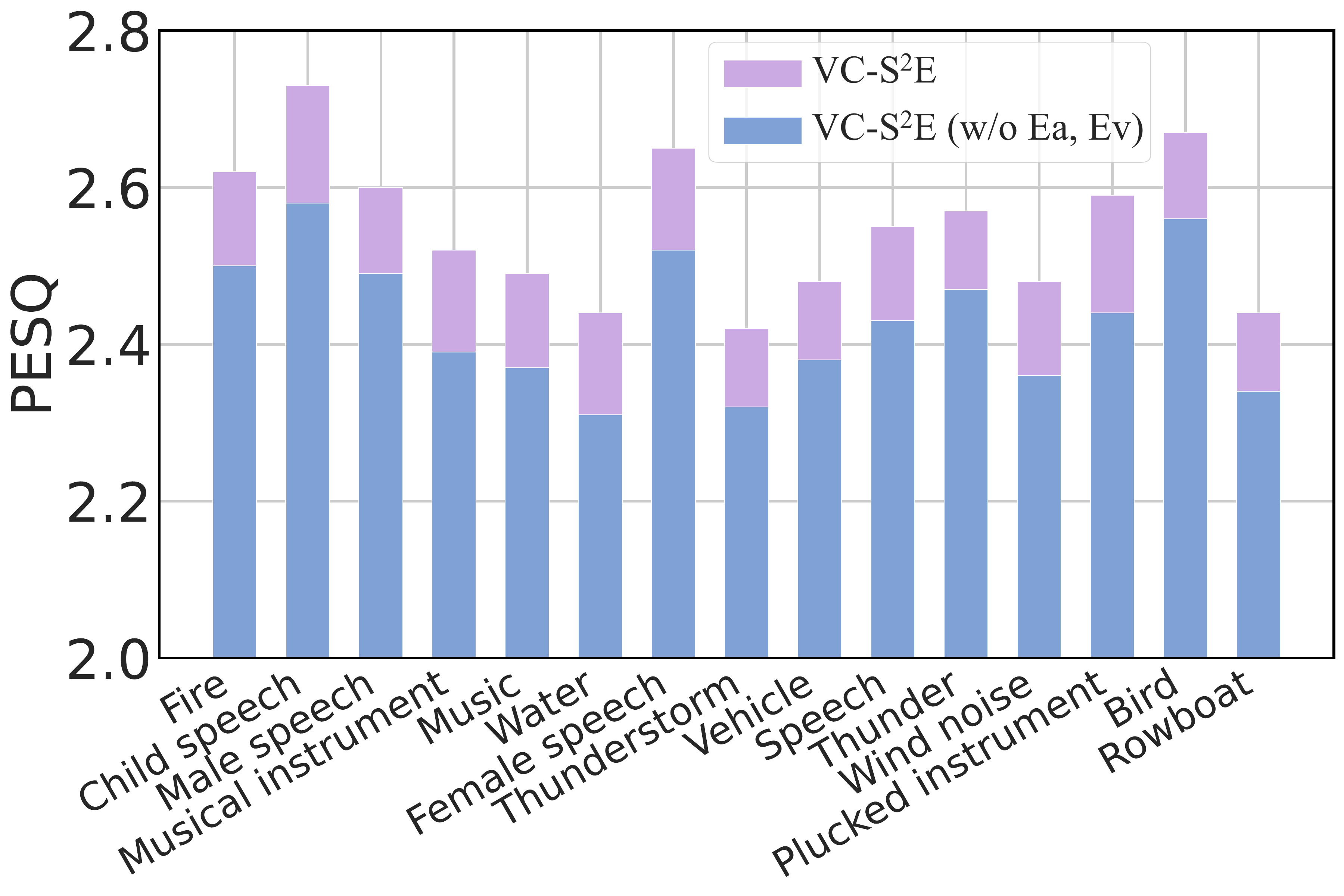}
    \subcaption{PESQ scores}  
    \label{fig2:1}
\end{subfigure}%
\hfill 
\begin{subfigure}[t]{.49\linewidth}
    \centering
    \includegraphics[width=\linewidth]{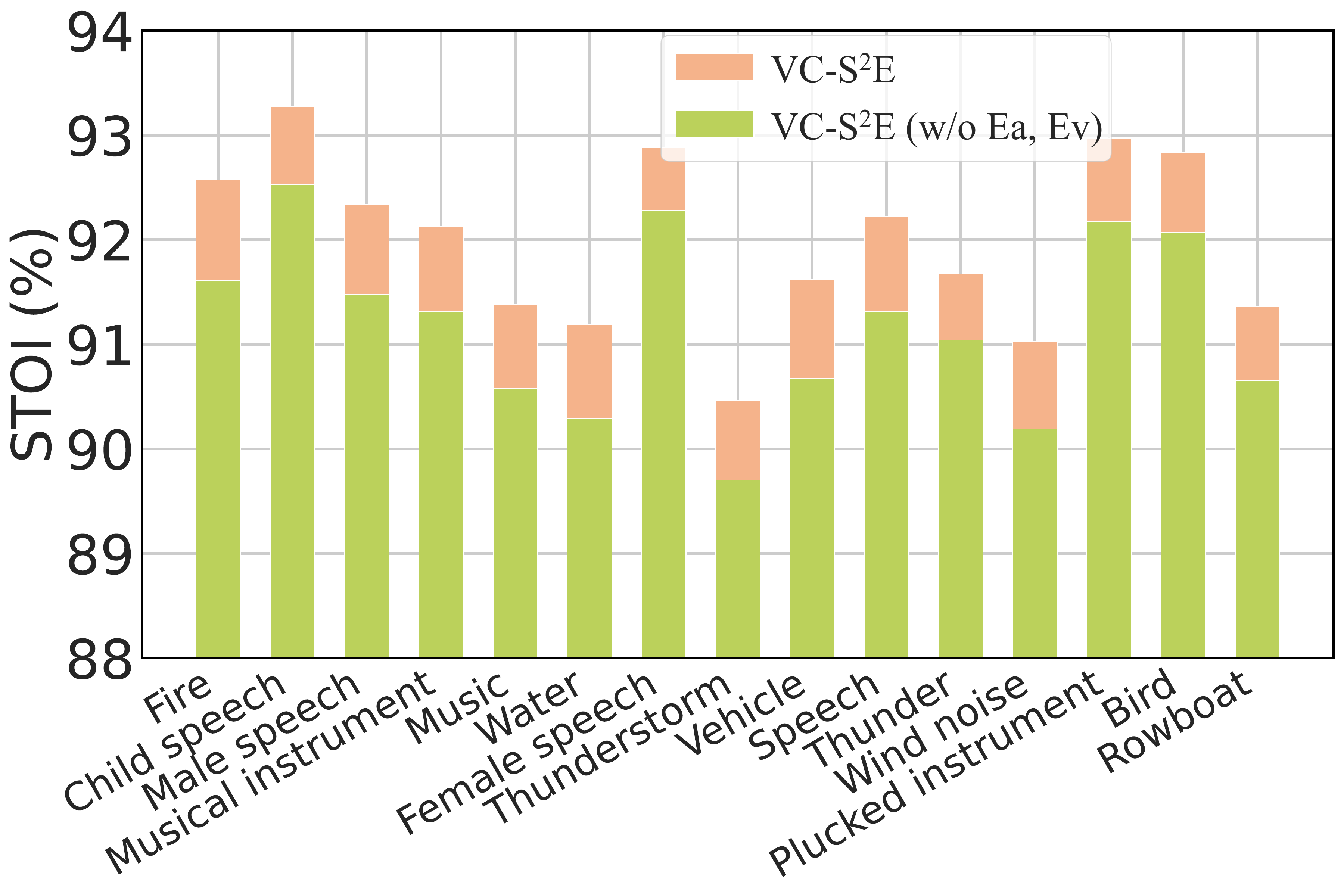}
    \subcaption{STOI scores}  
    \label{fig2:2}
\end{subfigure}
\caption{The (a) PESQ and (b) STOI scores for various audio categories when using AudioSet as the noise dataset. The results are averaged over all of the SNR conditions.}
\label{fig21}
\end{figure*}

\subsubsection{Implementation Details} Noise recordings exceeding $30$ seconds are segmented into clips no longer than 30 seconds, yielding $2\,500$ segments from MUSIC, $3\,130$ from AVSpeech, and $8\,560$ from AudioSet. For validation experiments, $1\,000$ clean speech utterances, $400$, $500$, and $1\,000$ noise recordings were randomly selected from the aforementioned speech and noise datasets, respectively. Each speech utterance is mixed with a random noise segment at a random \ac{SNR} sampled from $-20$ to $20$ dB (in 1 dB increments), to create the validation set. Clean speech data for testing are taken from the Librispeech \textit{test-clean-100} corpus. $200$ clean speech utterances are randomly picked (without replacement). $200$ noise recordings were randomly selected from the three AV datasets. Each clean speech recording is degraded by a random section from the noise recording at five SNR levels, i.e., -5 dB, 0 dB, 5 dB, 10 dB, and 15 dB. This produces $1\,000$ noisy AV clips for evaluation. All audio recordings are resampled at a rate of 16 kHz. The STFT spectrum is calculated using a Hanning window of 32 ms (512 samples) with a hop length of 16 ms (256 samples), resulting in a $257$-point magnitude spectrum as the acoustic input to the network. {For video processing, we follow CAV-MAE~\cite{gong2023contrastive} and employ a frame aggregation strategy to reduce computational overhead and fit our resources. In particular, we uniformly sample 10 RGB frames from each 10-second the video sequence (i.e., 1 FPS). At the training time, one RGB frame is randomly selected as the input. At the inference time, the visual embeddings of all sampled frames are averaged to generate the final scene visual embedding. Frame aggregation balances computational efficiency with the ability to capture
contextual visual information, ensuring meaningful complementary cues for the audio input.} In terms of time dimension, $T_y$ denotes the maximum time frame within a batch, while $T_a$ and $T_v$ are fixed at 128 and 49, respectively. The model dimension, $d_{model}$, is set to 256. For Mamba layers, we employ the default hyper-parameters~\cite{gu2023mamba}: the state dimension $d_{state} = 16$, convolution dimension $d_{conv} = 4$, and the expansion factor of 2.

\subsubsection{Training Strategy} The noisy mixtures are generated on the fly by dynamically mixing clean speech and noise clips at a random SNR sampled from $-10$ to $20$ dB (in 1 dB increments). All models are trained using a batch size of $10$ speech utterances for one gradient update. We employ the Adam optimizer with an initial learning rate of $1e^{-3}$, and default hyper-parameters, $\beta_{1}=0.9$, $\beta_{2}=0.999$. Following~\cite{mambaspeech}, we adopt the warm-up training strategy with $40\,000$ warm-up steps to adjust the learning rate. Gradient clipping is applied to clip gradients within the range of $-1$ to $1$~\cite{tfaj}. We use the mean square error (MSE) as the objective function. Training is carried out over a total of $150$ epochs. All experiments were carried out on an NVIDIA GeForce RTX 4090 GPU.

\subsubsection{Evaluation Metrics} We employ the \ac{PESQ}~\cite{pesq1} and the \ac{STOI}~\cite{estoi} as the evaluation metrics. 
Specifically, PESQ measures human-rated speech quality by predicting the Mean Opinion Score (MOS) for objective speech perception, ranging from -0.5 to 4.5, with higher values indicating better quality. STOI assesses speech intelligibility based on spectral correlation, typically ranging from 0 to 1, where higher scores reflect greater intelligibility.


\begin{figure*}[!tb]
    \centering
    \includegraphics[width=\linewidth]{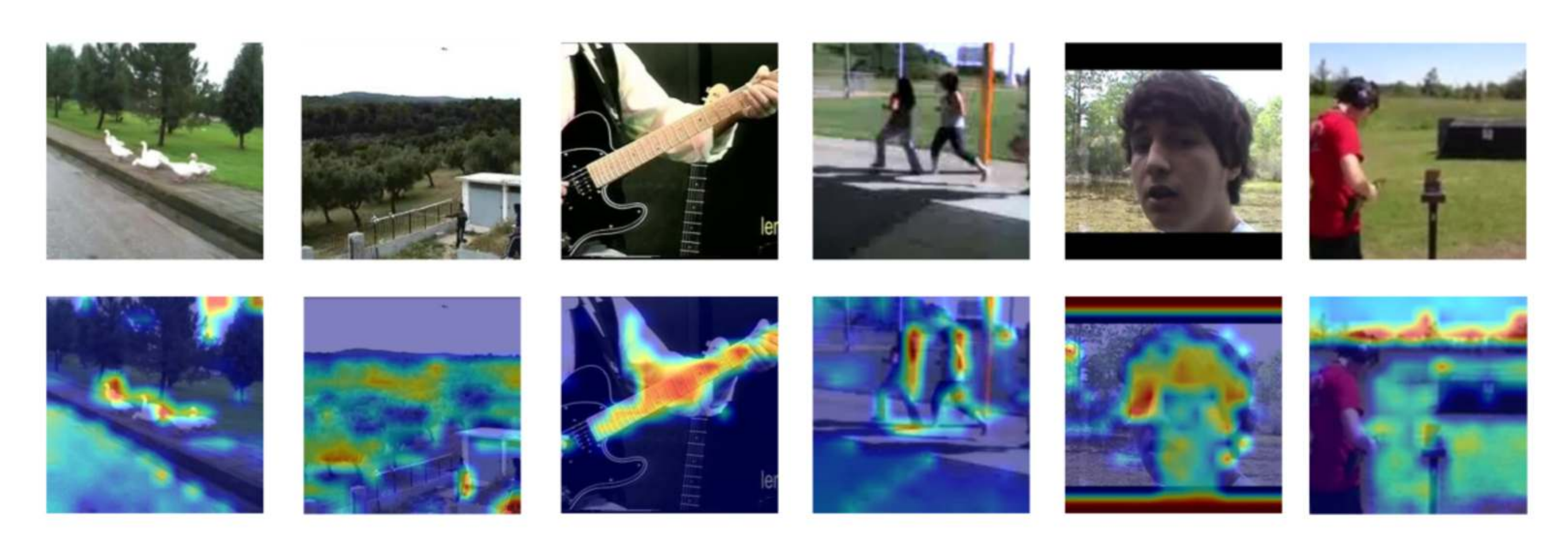}
    \caption{The Grad-CAM visualization of the focal areas in various video frames processed by our \ac{VSE} model. Top row: original video frames; bottom row: corresponding Grad-CAM heatmaps which emphasize regions crucial for detecting noise sources, as identified by the video encoder. The color varying from blue to red indicates the more significant regions.
    }
    \label{fig:gradcam}
\end{figure*}

\begin{table}[!t]
    \centering
    \small
      \caption{Ablation studies of the model inputs and architectures on  MUSIC noise dataset (${\bf E}_a$: scenario-aware audio embedding; ${\bf E}_v$: visual embedding; \textit{w/o} without).}
    \begin{tabular}{lcc}
    \toprule[1.0pt]
    \textbf{Ablations}              & \bf PESQ & \bf STOI(\%)         \\
        \midrule
    \ac{VSE} (ConMamba)                & 3.02     & 94.11                \\
        $ \ \ \ \ \ \ \  \ \ \ \  \ \ \  \ \  \ $ \textit{w/o} ${\bf E}_a$          & 2.81    & 93.25              \\
    $ \ \ \ \ \ \ \  \ \ \ \  \ \ \   \ \  \ $ \textit{w/o} ${\bf E}_a, {\bf E}_v$          & 2.77    & 93.09              \\
    \midrule
     $ \ \ \ \ \ \ \  \ \ \ \  \ \  \ $ (Conformer)                    & 2.73 & 92.81              \\
             $ \ \ \ \ \ \ \  \ \ \ \  \ \ \  \ \  \ $ \textit{w/o} ${\bf E}_a$          & 2.66  & 92.63            \\
    $ \ \ \ \ \ \ \  \ \ \ \  \ \ \   \ \  \ $ \textit{w/o} ${\bf E}_a, {\bf E}_v$          & 2.54   & 91.65              \\
    \toprule[1.0pt]
    \end{tabular}
            \label{tab:different_input}
    \vspace{-1.5em}
\end{table}

\subsection{Experimental Results} 
Table~\ref{tab:comparison_results} showcases the PESQ and STOI scores of the models across five \ac{SNR} levels, with the best results are highlighted in \textbf{bold}. The evaluation results demonstrate that our \ac{VSE} model consistently achieves superior performance in all three datasets. In the MUSIC dataset,  the unprocessed data have an average PESQ and STOI of 1.47 and 83.65\%, respectively. By applying conformer-based MP-SENet, the resulting enhanced speech has improved PESQ and STOI to 2.54 and 91.65\%, respectively. The AViTAR method, which employs a Conformer-based cross-attention AV framework, achieving a PESQ of 2.73 and a STOI of 92.81\%. \textcolor{black}{In contrast}, our \ac{VSE} proposal shows superior performance by incorporating the BiMamba architecture with an average PESQ of 3.02 and STOI of 94.11\%.
It can also be seen that by integrating the spectrum, visual, and scenario-aware audio embedding, \ac{VSE} exceeds ExtBiMamba with an average PESQ and STOI of 2.77 and 93.09\%, respectively.
{Moreover, the AVDCNN method leverages a dual-stream CNN with late audio-visual fusion, resulting in an average PESQ of 2.23 and STOI of 89.83\%, respectively. MLDM, which utilizes a multi-level distortion measure objective, achieves an average PESQ of 2.51 and STOI of 91.22\%. Meanwhile, ALTKD employs a U-Net-based teacher-student framework that integrates ultrasound tongue images and lip videos; however, it attains only an average PESQ of 2.05 and STOI of 87.83\%. Notably, the DNE method, which exploits VAD to extract noise features for robust SE, records an average PESQ of 2.93 and STOI of 93.41\%. Despite the competitive performance of DNE at specific SNR levels, our proposal still outperforms all these approaches, achieving an average PESQ of 3.02 and STOI of 94.11\%.
}
Furthermore, the same observations can be seen on the AVSpeech and AudioSet datasets as well, where our proposed \ac{VSE} achieves superior performance with the PESQ and STOI of 2.76 and 92.70\% for AVSpeech, 2.59 and 92.25\% for AudioSet, respectively. 

\begin{figure}[!tb]
\centering
\begin{subfigure}[t]{.49\linewidth}
    \centering
    \includegraphics[width=\linewidth]{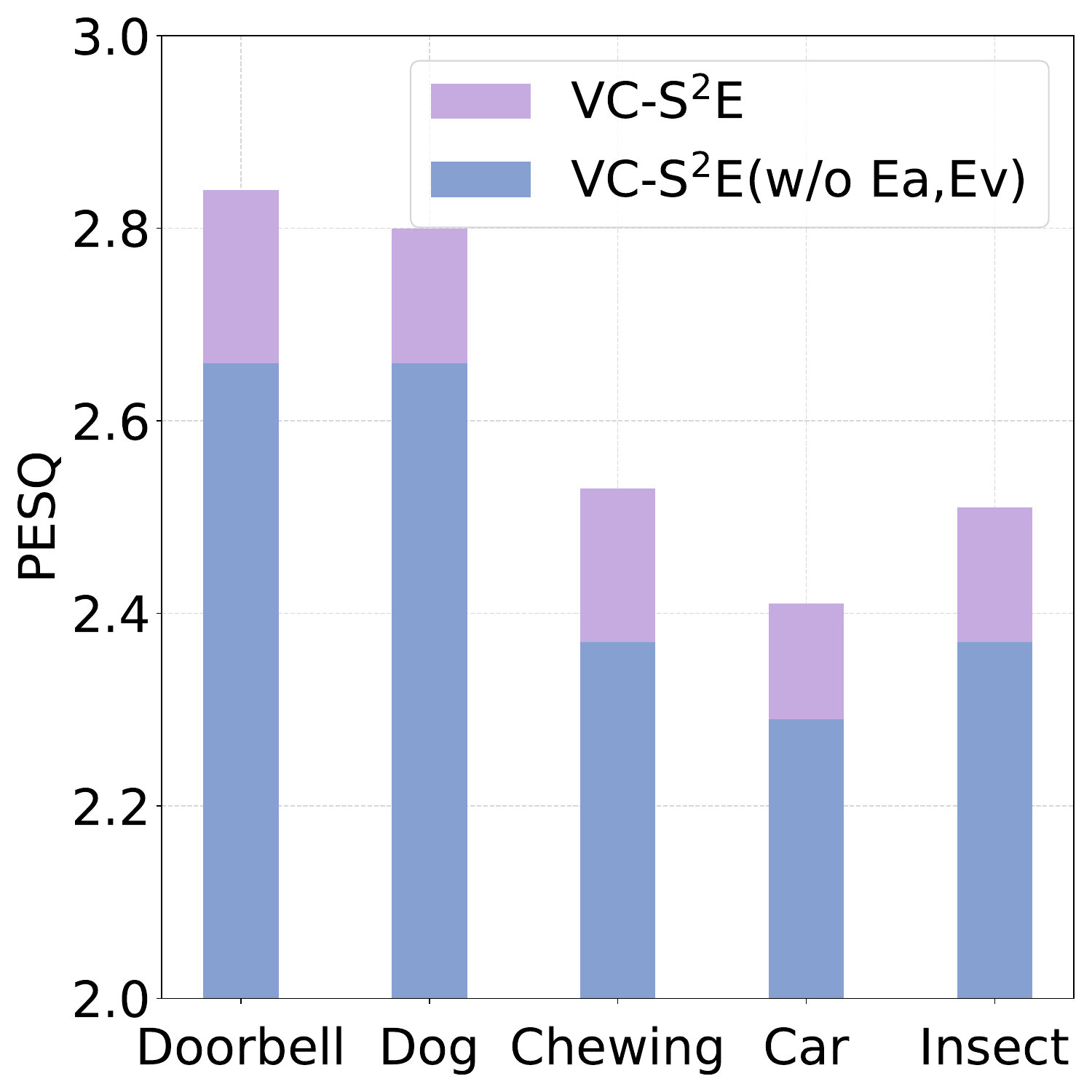}
    \subcaption{PESQ scores}  
    \label{fig:PESQunseen}
\end{subfigure}%
\begin{subfigure}[t]{.49\linewidth}
    \centering
    \includegraphics[width=\linewidth]{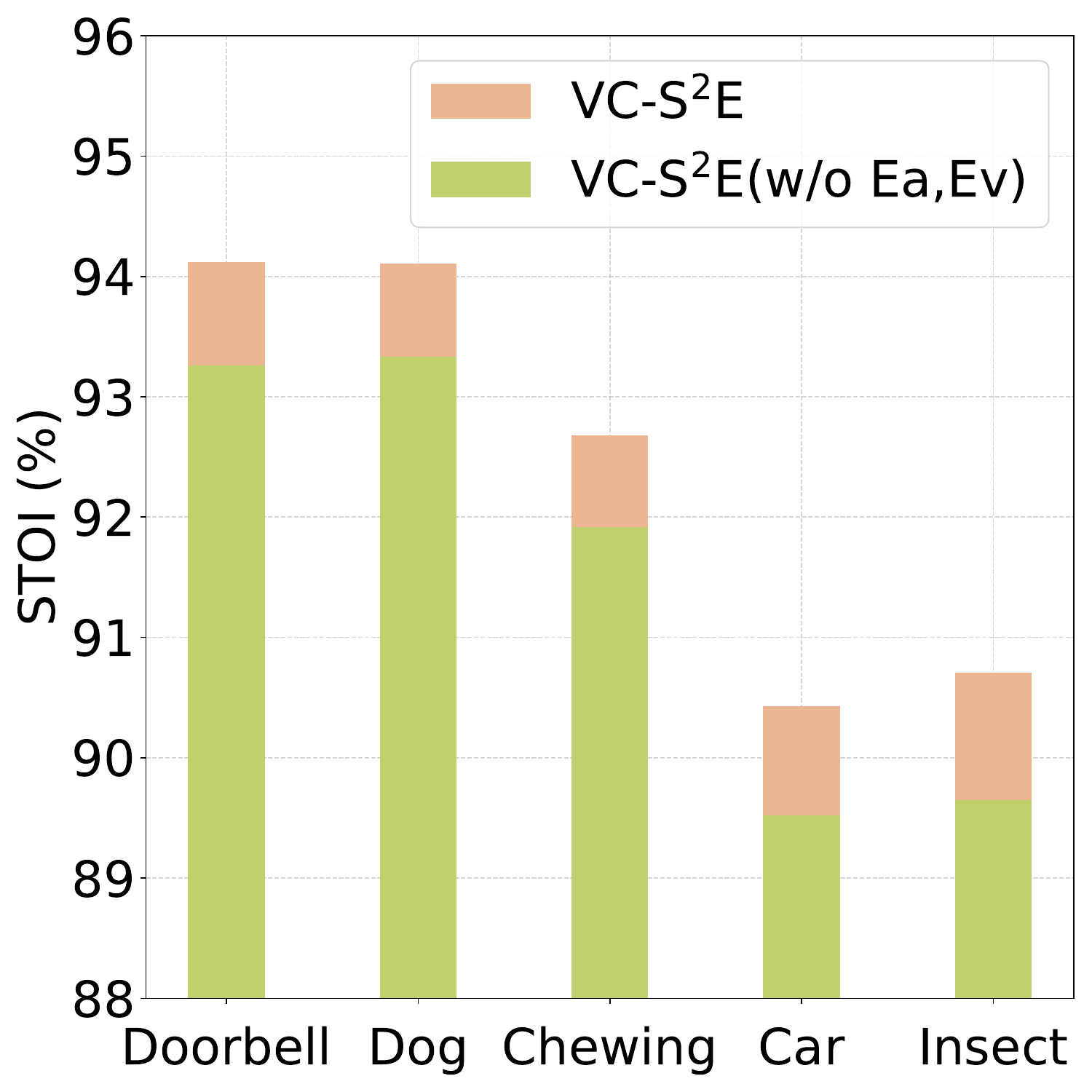}
    \subcaption{STOI scores}  
    \label{fig:STOIunseen}
\end{subfigure}
\caption{The (a) PESQ and (b) STOI for scenes unseen at the training time.}
\label{fig:unseen}
\end{figure}

Figure~\ref{fig21} illustrates the PESQ and STOI of our proposed \ac{VSE} method and its counterpart
without visual and scenario-aware embeddings where AudioSet is used as the noise dataset.
In both figures, \ac{VSE} consistently outperforms the other
across all categories. 
For example, the PESQ of `Fire' improves from 2.50 to 2.62 while for category `Child speech', the score improves from 2.58 to 2.73.
Similarly, the STOI scores (Figure~\ref{fig2:2}) also show significant improvements.
For `Fire', the STOI score increases from 91.61\% to 92.57\% while for `Child speech', it improves from 92.53\% to 93.27\%.
These results clearly reveal that incorporating visual and audio scenario-aware embeddings leads to substantial improvements in both perceptual quality and intelligibility across each individual audio category.
Furthermore, we also test the model performance on five new categories (i.e., dog, insect, chewing, car and doorbell) where the results indicated in Figure~\ref{fig:unseen} show the generalization of our proposed model.


\begin{figure*}[!t]
    \centering
    \includegraphics[width=\linewidth]{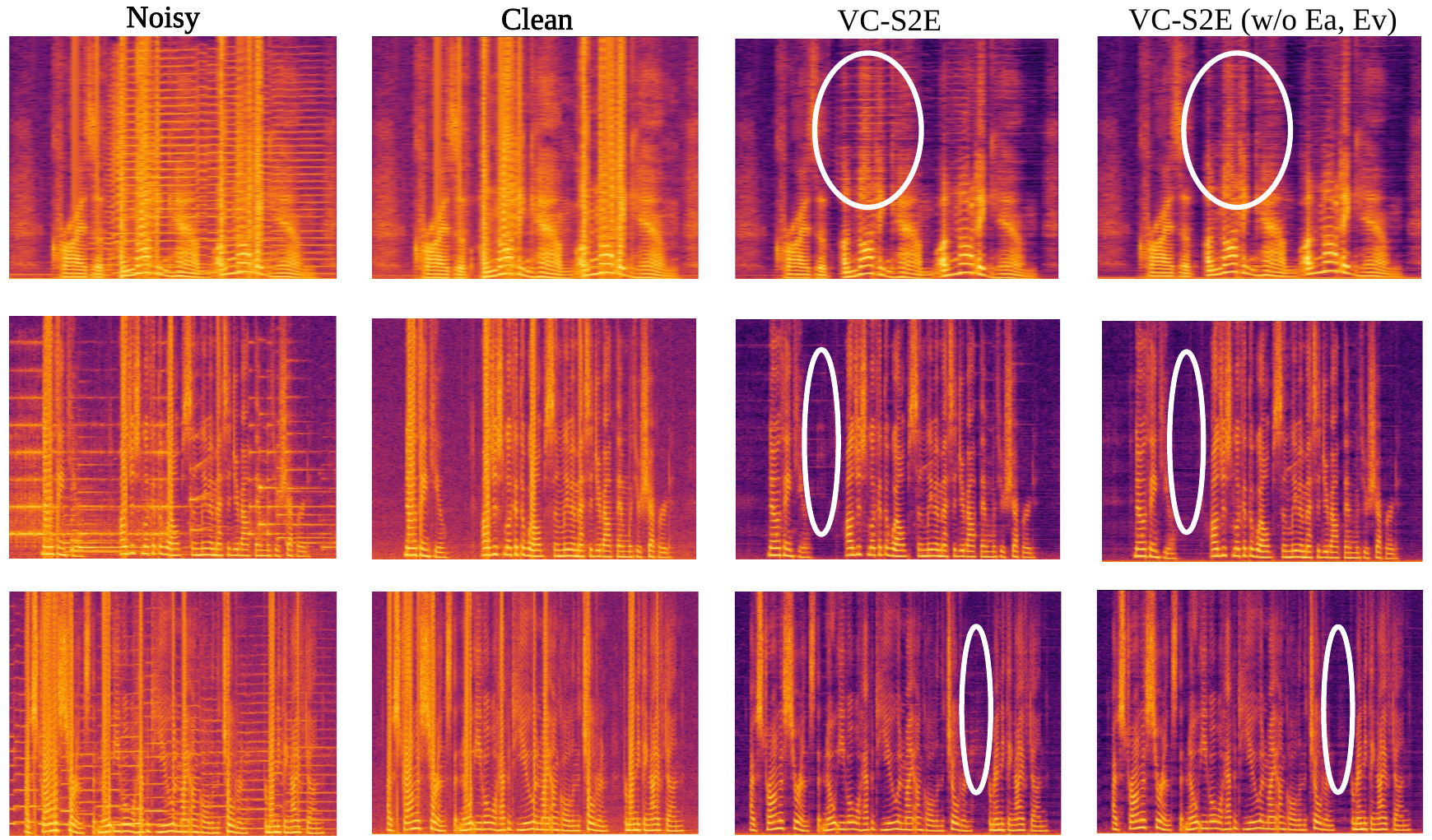}
    \caption{The visualization of log-magnitude spectrograms of noisy, clean, and enhanced speech signals produced by \ac{VSE} and \ac{VSE} (w/o $\mathbf{E}_{a}$, $\mathbf{E}_{v}$). 
    }
    \label{fig:STFT}
    \vspace{-1.5em}
\end{figure*}
\subsection{Ablation Study}
\subsubsection{Model Input and Architecture}
In Table \ref{tab:different_input}, we ablate the model inputs and architectures using the MUSIC noise dataset.
When removing the scenario-aware audio embedding $\mathbf{E}_a$ from our proposed \ac{VSE} model, the PESQ and STOI values were reduced by 0.21 and 0.86\%.
Moreover, without the video prior $\mathbf{E}_v$, PESQ and STOI further degrade to 2.77 and 93.09\%, respectively.
This is because $\mathbf{E}_a$ and $\mathbf{E}_v$ are pre-trained through contrastive learning, adding $\mathbf{E}_a$ can better integrate visual embedding $\mathbf{E}_v$ in the subsequent fusion part. 
The \textcolor{black}{same trends are observed} when using Conformer as the backbone. For example, removing scenario-aware audio embedding $\mathbf{E}_a$ resulted in the degraded PESQ and STOI from 2.73 and 92.81\% to 2.66 and 92.63\%.  Further removal of $\mathbf{E}_v$ provides the worst results with the PESQ and STOI of 2.54 and 91.65\%, respectively.


\begin{table*}[!hbtp]
    \centering
    \small
    \caption{\textcolor{black}{Ablation studies of different feature combination strategies on MUSIC noise dataset ($\oplus$: element-wise addition; $\concat$: concatenation; $\odot$: element-wise multiplication).} }
    \scalebox{1.0}{
    \begin{tabular}{@{}l ccccc c ccccc c@{}}
    \toprule[1.0pt]
    \multirow{2}{*}{\textbf{Fusion Strategy}} & \multicolumn{6}{c}{\textbf{PESQ}} & \multicolumn{6}{c}{\textbf{STOI} (in \%)} \\
    \cmidrule(lr){2-13}
    & \bf -5 dB & \bf 0 dB & \bf 5 dB & \bf 10 dB & \bf 15 dB & \bf Avg. & \bf -5 dB & \bf 0 dB & \bf 5 dB & \bf 10 dB & \bf 15 dB & \bf Avg. \\
    \midrule
    $\mathbf{E}_{y}$                                                            & 1.97 & 2.39 & 2.82 & 3.13 & 3.54 & 2.77 & 87.56 & 91.34 & 94.09 & 95.65 & 96.82 & 93.09 \\ 
    $\mathbf{E}_{y}$ $\concat$ \ ($\mathbf{E}_{a}$ $\oplus$ \ $\mathbf{E}_{v}$) & 2.04 & 2.46 & 2.90 & 3.21 & 3.57 & 2.84 & 88.03 & 91.62 & 94.26 & 95.72 & 96.75 & 93.28 \\ 
    $\mathbf{E}_{y}$ $\concat$ \ ($\mathbf{E}_{a}$ $\odot$ \ $\mathbf{E}_{v}$)  & 2.01 & 2.43 & 2.86 & 3.19 & 3.53 & 2.80 & 87.58 & 91.25 & 94.11 & 95.62 & 96.56 & 93.02 \\ 
    \midrule
    \textbf{\ac{VSE}(our)}                                                      & \textbf{2.23} & \textbf{2.66} & \textbf{3.08} & \textbf{3.38} & \textbf{3.73} & \textbf{3.02} & \textbf{89.55} & \textbf{92.61} & \textbf{94.96} & \textbf{96.24} & \textbf{97.18} & \textbf{94.11} \\
    \toprule[1.0pt]
    \end{tabular}
    }
    \label{tab:ablation_study}
    \vspace{-1.0em}
\end{table*}

\subsubsection{Feature Combination}
In Table \ref{tab:ablation_study}, we evaluate various feature combination strategies.
Initially, we present the model's performance using only the spectral embedding $\mathbf{E}_y$, resulting in a PESQ of 2.77 and a STOI of 93.09\%.
We then analyze different feature combination strategies. Using $\mathbf{E}_y$ concatenated with ($\mathbf{E}_a$ $\oplus$\ $\mathbf{E}_v$) yields a slight improvement, achieving a PESQ of 2.84 and a STOI of 93.28\%. Similarly, the strategy of $\mathbf{E}_y$ concatenated with ($\mathbf{E}_a$ $\odot$\ $\mathbf{E}_v$) shows some improvement over the baseline, but not substantially.
The best performance is achieved with our employed fusion strategy: $\mathbf{E}_y \ \concat\ \mathbf{E}_a \ \concat\ \mathbf{E}_v$, resulting in a PESQ of 3.02 and a STOI of 94.11\%. This strategy not only offers superior performance but also maintains the simplest complexity.

\subsubsection{Visualization}
We employ \ac{GCAM}, a widely used scheme, to visually interpret which image regions have significant contributions to SE. As illustrated in Figure~\ref{fig:gradcam},  the original video frames are shown in the upper row
while the corresponding Grad-CAM heatmaps are depicted  the bottom row, which emphasize regions crucial for detecting noise sources. The colors varying from blue to red correspond to the higher values assigned.
These heatmaps reveal the model's ability to focus on relevant contextual elements, such as moving ducks, pedestrians, and the guitar, etc. which depicts the origins of the noise 
and indicates that the detected visual sounding object assists SE. 


In Figure~\ref{fig:STFT}, we present a comparison of different log-magnitude spectrograms, showcasing the results for noisy audio, clean speech, and the speech generated by our proposed \ac{VSE} model under two different configurations: without (\textit{w/o}) and with (\textit{w}) the scenario-aware audio embedding ${\bf E}_a$ and visual embedding ${\bf E}_v$. From left to right, the images represent the spectrograms of noisy audio, clean speech, and the outputs of our \ac{VSE} model under the two conditions.
Notably, when comparing the clean speech spectrograms with those generated by our model, we observe that the speech generated with scenario-aware embeddings exhibits a clearer and more accurate representation of the target speech signal. 
This improvement is highlighted by the white-circled areas in the spectrograms, where the noise reduction is most evident.
These observations are consistent with the quantitative results presented in Table~\ref{tab:different_input}, where we report the performance of our model in terms of PESQ  and STOI.
Thus, the results from the spectrogram analysis and the objective evaluation confirm that our proposal achieves significant SE improvements, leveraging the complementary nature of visual scene information.

\begin{table}[!tb]
\centering
\footnotesize
\def\arraystretch{1.3}  
\setlength{\tabcolsep}{10pt}  
\caption{Experimental results using the VoxBlink dataset (with co-speech facial motions) as the target clean speech.}
\scalebox{1}{
\begin{tabular}{@{}c l c c@{}}
\toprule[1.0pt]
\textbf{Dataset} & \textbf{Model}  & \textbf{PESQ} & \textbf{STOI (in \%)} \\
\midrule
\multirow{4}{*}{\textbf{MUSIC}} 
& ExtBiMamba                   & 2.28 & 86.50 \\ 
& \textbf{\ac{VSE}(our)}       & 2.52 & 88.05 \\
& ——(only Lip)                 & 2.63 & 88.69 \\
& ——(Lip+Scene)                & \textbf{2.71} & \textbf{89.16} \\
\midrule
\multirow{4}{*}{\textbf{AVSpeech}} 
& ExtBiMamba                   & 2.33 & 85.86 \\ 
& \textbf{\ac{VSE}(our)}       & 2.50 & 87.29 \\
& ——(only Lip)                 & 2.58 & 87.83 \\
& ——(Lip+Scene)                & \textbf{2.63} & \textbf{88.24} \\
\midrule
\multirow{4}{*}{\textbf{AudioSet}} 
& ExtBiMamba                   & 2.01 & 80.84 \\
& \textbf{\ac{VSE}(our)}       & 2.20 & 82.64 \\
& ——(only Lip)                 & 2.31 & 83.82 \\
& ——(Lip+Scene)                & \textbf{2.33} & \textbf{84.19} \\
\toprule[1.0pt]
\end{tabular}}
\label{VoxBlink_results}
\end{table}

\subsection{Discussion}
To better evaluate the SAV-SE framework, we conducted new experiments using the VoxBlink \cite{lin2024VoxBlink} dataset as the target clean speech. 
This dataset inherently includes facial videos, from which detailed lip information can be extracted. 
For comparison, we considered the following setups:
1) ExtBiMamba: adopts the Mamba architecture and relies solely on pure audio features;
2) \ac{VSE}(our): utilizes the visual background cue;
3) only Lip: uses only lip features extracted from the target speaker's visual stream, and the Lip+Scene: integrates both the lip features and the background visual cues.

In the MUSIC dataset, \ac{VSE}(our) achieved an average PESQ of 2.52 and an average STOI of 88.05\%, whereas the only Lip method reached 2.63 and 88.69\%, respectively.
This may be because lip motion features inherently capture finer temporal coherence related to speech, which can enhance SE performance to some extent.
Moreover, when exploring both the lip features and visual scene cues in the Lip+Scene method, its average PESQ and STOI scores further increased to 2.71 and 89.16\%, respectively, indicating the contribution from both lip and the scene cues. Notably, a similar trend was also observed in the AVSpeech and AudioSet datasets, demonstrating that fusing multiple visual cues exhibits a potential improvement.
    
\FloatBarrier
\section{Conclusion and Future Work}
In this paper, we introduce a novel \ac{SAVSE} task that addresses speech enhancement by leveraging visual information from noisy environments. In contrast to existing works that rely on visual lip or face motions, our work is the first to enhance speech using rich contextual information from synchronized video as an auxiliary cue. Specifically, our proposed \ac{VSE} method integrates the Conformer and Mamba modules, using their distinct advantages to create a robust framework that distinguishes and mitigates environmental noise. Extensive experiments conducted on three different datasets demonstrate the superiority of \ac{VSE} over the other competitive methods, achieving the improved PESQ of 3.64, 3.48, 3.38, and STOI of 93.52\%, 91.25\%, 92.33\% on MUSIC, AVSpeech and AudioSet, respectively.

Despite effective, our proposal still has several limitations. Currently, it focuses exclusively on speech enhancement. In future work, we plan to extend our approach to tackle speech separation, enabling the model to distinguish and enhance each individual speaker in complex acoustic environments. Moreover, real-world applications often involve incomplete or occluded data, such as partially missing speech signals, occluded visual scenes, or obscured lip regions. To improve SAV-SE’s robustness, we will explore strategies to ensure reliable performance even when key modalities are partially missing or degraded.



\bibliographystyle{IEEEtran} 
\bibliography{ref}

\begin{thebibliography}{10}
\providecommand{\url}[1]{#1}
\csname url@samestyle\endcsname
\providecommand{\newblock}{\relax}
\providecommand{\bibinfo}[2]{#2}
\providecommand{\BIBentrySTDinterwordspacing}{\spaceskip=0pt\relax}
\providecommand{\BIBentryALTinterwordstretchfactor}{4}
\providecommand{\BIBentryALTinterwordspacing}{\spaceskip=\fontdimen2\font plus
\BIBentryALTinterwordstretchfactor\fontdimen3\font minus \fontdimen4\font\relax}
\providecommand{\BIBforeignlanguage}[2]{{%
\expandafter\ifx\csname l@#1\endcsname\relax
\typeout{** WARNING: IEEEtran.bst: No hyphenation pattern has been}%
\typeout{** loaded for the language `#1'. Using the pattern for}%
\typeout{** the default language instead.}%
\else
\language=\csname l@#1\endcsname
\fi
#2}}
\providecommand{\BIBdecl}{\relax}
\BIBdecl

\bibitem{loizou}
P.~C. Loizou, \emph{Speech Enhancement: Theory and Practice}, 2nd~ed.\hskip 1em plus 0.5em minus 0.4em\relax Boca Raton, FL, USA: CRC Press, Inc., 2013.

\bibitem{overview2018}
D.~Wang and J.~Chen, ``Supervised speech separation based on deep learning: An overview,'' \emph{IEEE/ACM Trans. on Audio, Speech and Language Processing}, vol.~26, no.~10, pp. 1702--1726, 2018.

\bibitem{borgstrom2010improved}
B.~J. Borgstr{\"o}m and A.~Alwan, ``Improved speech presence probabilities using hmm-based inference, with applications to speech enhancement and asr,'' \emph{IEEE Journal of Selected Topics in Signal Processing}, vol.~4, no.~5, pp. 808--815, 2010.

\bibitem{xian2020multi}
Y.~Xian, Y.~Sun, W.~Wang, and S.~M. Naqvi, ``A multi-scale feature recalibration network for end-to-end single channel speech enhancement,'' \emph{IEEE Journal of Selected Topics in Signal Processing}, vol.~15, no.~1, pp. 143--155, 2020.

\bibitem{gulati2020conformer}
A.~Gulati, J.~Qin, C.-C. Chiu, N.~Parmar, Y.~Zhang, J.~Yu, W.~Han, S.~Wang, Z.~Zhang, Y.~Wu \emph{et~al.}, ``Conformer: Convolution-augmented transformer for speech recognition,'' \emph{arXiv preprint arXiv:2005.08100}, 2020.

\bibitem{wang2022predict}
J.~Wang, X.~Qian, and H.~Li, ``Predict-and-update network: Audio-visual speech recognition inspired by human speech perception,'' \emph{arXiv preprint arXiv:2209.01768}, 2022.

\bibitem{rodomagoulakis2019improved}
I.~Rodomagoulakis and P.~Maragos, ``Improved frequency modulation features for multichannel distant speech recognition,'' \emph{IEEE Journal of Selected Topics in Signal Processing}, vol.~13, no.~4, pp. 841--849, 2019.

\bibitem{liu2024aligning}
H.~Liu, X.~Zhang, H.~Zhang, L.~P. Garcia, A.~W. Khong, E.~S. Chng, and S.~Watanabe, ``Aligning speech to languages to enhance code-switching speech recognition,'' \emph{arXiv preprint arXiv:2403.05887}, 2024.

\bibitem{wan2018generalized}
L.~Wan, Q.~Wang, A.~Papir, and I.~L. Moreno, ``Generalized end-to-end loss for speaker verification,'' in \emph{Proc. of {IEEE} Int. Conf. on Audio, Speech and Signal Processing}, 2018, pp. 4879--4883.

\bibitem{qian2022audio}
X.~Qian, Z.~Wang, J.~Wang, G.~Guan, and H.~Li, ``Audio-visual cross-attention network for robotic speaker tracking,'' \emph{IEEE/ACM Trans. on Audio, Speech and Language Processing}, vol.~31, pp. 550--562, 2022.

\bibitem{qian2021audio}
X.~Qian, A.~Brutti, O.~Lanz, M.~Omologo, and A.~Cavallaro, ``Audio-visual tracking of concurrent speakers,'' \emph{IEEE Trans. on Multimedia}, 2021.

\bibitem{qian2021multi}
X.~Qian, M.~Madhavi, Z.~Pan, J.~Wang, and H.~Li, ``Multi-target {DOA} estimation with an audio-visual fusion mechanism,'' in \emph{Proc. of {IEEE} Int. Conf. on Audio, Speech and Signal Processing}, 2021, pp. 2814--2818.

\bibitem{mmse}
Y.~Ephraim and D.~Malah, ``{Speech Enhancement Using a Minimum Mean-Square Error Short-Time Spectral Amplitude Estimator},'' \emph{IEEE/ACM Trans. on Audio, Speech and Language Processing}, vol. ASSP-32, no.~6, pp. 1109--1121, Dec. 1984.

\bibitem{2007mmse}
J.~S. Erkelens, R.~C. Hendriks, R.~Heusdens, and J.~Jensen, ``Minimum mean-square error estimation of discrete fourier coefficients with generalized gamma priors,'' \emph{IEEE/ACM Trans. on Audio, Speech and Language Processing}, vol.~15, no.~6, pp. 1741--1752, 2007.

\bibitem{zhang2019}
Q.~Zhang, M.~Wang, Y.~Lu, L.~Zhang, and M.~Idrees, ``A novel fast nonstationary noise tracking approach based on mmse spectral power estimator,'' \emph{Digital Signal Processing}, vol.~88, pp. 41--52, 2019.

\bibitem{LogNC}
Q.~Zhang, M.~Wang, Y.~Lu, M.~Idrees, and L.~Zhang, ``Fast nonstationary noise tracking based on log-spectral power mmse estimator and temporal recursive averaging,'' \emph{IEEE Access}, vol.~7, pp. 80\,985--80\,999, 2019.

\bibitem{sivaraman2022efficient}
A.~Sivaraman and M.~Kim, ``Efficient personalized speech enhancement through self-supervised learning,'' \emph{IEEE Journal of Selected Topics in Signal Processing}, vol.~16, no.~6, pp. 1342--1356, 2022.

\bibitem{taha2018survey}
T.~M. Taha, A.~Adeel, and A.~Hussain, ``A survey on techniques for enhancing speech,'' \emph{International Journal of Computer Applications}, vol. 179, no.~17, pp. 1--14, 2018.

\bibitem{mambaspeech}
X.~Zhang, Q.~Zhang, H.~Liu, T.~Xiao, X.~Qian, B.~Ahmed, E.~Ambikairajah, H.~Li, and J.~Epps, ``Mamba in speech: Towards an alternative to self-attention,'' \emph{arXiv}, 2024.

\bibitem{wei2022learning}
Y.~Wei, D.~Hu, Y.~Tian, and X.~Li, ``Learning in audio-visual context: A review, analysis, and new perspective,'' \emph{arXiv preprint arXiv:2208.09579}, 2022.

\bibitem{lin2021exploiting}
Y.-B. Lin and Y.-C.~F. Wang, ``Exploiting audio-visual consistency with partial supervision for spatial audio generation,'' in \emph{Proceedings of the AAAI Conference on Artificial Intelligence}, vol.~35, no.~3, 2021, pp. 2056--2063.

\bibitem{gabbay2017visual}
A.~Gabbay, A.~Shamir, and S.~Peleg, ``Visual speech enhancement,'' \emph{arXiv preprint arXiv:1711.08789}, 2017.

\bibitem{afouras2018conversation}
T.~Afouras, J.~S. Chung, and A.~Zisserman, ``The conversation: Deep audio-visual speech enhancement,'' \emph{arXiv preprint arXiv:1804.04121}, 2018.

\bibitem{gabbay2018seeing}
A.~Gabbay, A.~Ephrat, T.~Halperin, and S.~Peleg, ``Seeing through noise: Visually driven speaker separation and enhancement,'' in \emph{Proc. of {IEEE} Int. Conf. on Audio, Speech and Signal Processing}.\hskip 1em plus 0.5em minus 0.4em\relax IEEE, 2018, pp. 3051--3055.

\bibitem{wiener1996}
P.~{Scalart} and J.~V. {Filho}, ``Speech enhancement based on a priori signal to noise estimation,'' in \emph{Proc. of {IEEE} Int. Conf. on Audio, Speech and Signal Processing}, vol.~2, 1996, pp. 629--632.

\bibitem{PALIWAL2010450}
K.~Paliwal, K.~Wójcicki, and B.~Schwerin, ``Single-channel speech enhancement using spectral subtraction in the short-time modulation domain,'' \emph{Speech Communication}, vol.~52, no.~5, pp. 450--475, 2010.

\bibitem{mmse2017}
M.~Krawczyk-Becker and T.~Gerkmann, ``On {MMSE}-based estimation of amplitude and complex speech spectral coefficients under phase-uncertainty,'' \emph{IEEE/ACM Trans. on Audio, Speech and Language Processing}, vol.~24, no.~12, pp. 2251--2262, 2016.

\bibitem{IMM2006-04511}
M.~N. Schmidt and R.~K. Olsson, ``Single-channel speech separation using sparse non-negative matrix factorization,'' sep 2006.

\bibitem{purwins2019deep}
H.~Purwins, B.~Li, T.~Virtanen, J.~Schluter, S.-Y. Chang, and T.~Sainath, ``Deep learning for audio signal processing,'' \emph{IEEE Journal of Selected Topics in Signal Processing}, vol.~13, no.~2, pp. 206--219, 2019.

\bibitem{cleanunet}
Z.~Kong, W.~Ping, A.~Dantrey, and B.~Catanzaro, ``Speech denoising in the waveform domain with self-attention,'' in \emph{Proc. of {IEEE} Int. Conf. on Audio, Speech and Signal Processing}, 2022, pp. 7867--7871.

\bibitem{demcus}
A.~Defossez, G.~Synnaeve, and Y.~Adi, ``Real time speech enhancement in the waveform domain,'' in \emph{Proc. Interspeech}, 2020.

\bibitem{kolbaek2020loss}
M.~Kolb{\ae}k, Z.-H. Tan, S.~H. Jensen, and J.~Jensen, ``On loss functions for supervised monaural time-domain speech enhancement,'' \emph{IEEE/ACM Trans. on Audio, Speech and Language Processing}, vol.~28, pp. 825--838, 2020.

\bibitem{tfaj}
Q.~Zhang, X.~Qian, Z.~Ni, A.~Nicolson, E.~Ambikairajah, and H.~Li, ``A time-frequency attention module for neural speech enhancement,'' \emph{IEEE/ACM Trans. on Audio, Speech and Language Processing}, vol.~31, pp. 462--475, 2023.

\bibitem{yongxu2015}
Y.~Xu, J.~Du, L.-R. Dai, and C.-H. Lee, ``A regression approach to speech enhancement based on deep neural networks,'' \emph{IEEE/ACM Trans. on Audio, Speech and Language Processing}, vol.~23, no.~1, pp. 7--19, 2014.

\bibitem{8910352}
K.~Tan and D.~Wang, ``Learning complex spectral mapping with gated convolutional recurrent networks for monaural speech enhancement,'' \emph{IEEE/ACM Trans. on Audio, Speech and Language Processing}, vol.~28, pp. 380--390, 2020.

\bibitem{chenlstm}
J.~Chen and D.~Wang, ``Long short-term memory for speaker generalization in supervised speech separation,'' \emph{The Journal of the Acoustical Society of America}, vol. 141, no.~6, pp. 4705--4714, 2017.

\bibitem{GRN}
K.~Tan, J.~Chen, and D.~Wang, ``Gated residual networks with dilated convolutions for monaural speech enhancement,'' \emph{IEEE/ACM Trans. on Audio, Speech and Language Processing}, vol.~27, no.~1, pp. 189--198, 2018.

\bibitem{DeepMMSE}
Q.~Zhang, A.~Nicolson, M.~Wang, K.~K. Paliwal, and C.~Wang, ``{DeepMMSE}: A deep learning approach to mmse-based noise power spectral density estimation,'' \emph{IEEE/ACM Trans. on Audio, Speech and Language Processing}, vol.~28, pp. 1404--1415, Jun. 2020.

\bibitem{zhang2024empirical}
Q.~Zhang, M.~Ge, H.~Zhu, E.~Ambikairajah, Q.~Song, Z.~Ni, and H.~Li, ``An empirical study on the impact of positional encoding in transformer-based monaural speech enhancement,'' in \emph{Proc. of {IEEE} Int. Conf. on Audio, Speech and Signal Processing}, 2024, pp. 1001--1005.

\bibitem{conformer-se}
E.~Kim and H.~Seo, ``Se-conformer: Time-domain speech enhancement using conformer,'' in \emph{Proc. Interspeech 2021}, 2021, pp. 2736--2740.

\bibitem{Lee2020DynamicNE}
J.~Lee, Y.~Jung, M.~Jung, and H.~Kim, ``Dynamic noise embedding: Noise aware training and adaptation for speech enhancement,'' \emph{2020 Asia-Pacific Signal and Information Processing Association Annual Summit and Conference (APSIPA ASC)}, pp. 739--746, 2020.

\bibitem{ting2022speechenhancementbasedcyclegan}
W.-Y. Ting, S.-S. Wang, H.-L. Chang, B.~Su, and Y.~Tsao, ``Speech enhancement based on cyclegan with noise-informed training,'' 2022.

\bibitem{xiang2023twostagedeeprepresentationlearningbased}
Y.~Xiang, J.~L. Højvang, M.~H. Rasmussen, and M.~G. Christensen, ``A two-stage deep representation learning-based speech enhancement method using variational autoencoder and adversarial training,'' 2023.

\bibitem{gu2023mamba}
A.~Gu and T.~Dao, ``Mamba: Linear-time sequence modeling with selective state spaces,'' \emph{arXiv preprint arXiv:2312.00752}, 2023.

\bibitem{chao2024investigation}
R.~Chao, W.-H. Cheng, M.~La~Quatra, S.~M. Siniscalchi, C.-H.~H. Yang, S.-W. Fu, and Y.~Tsao, ``An investigation of incorporating mamba for speech enhancement,'' \emph{arXiv preprint arXiv:2405.06573}, 2024.

\bibitem{jiang2024dual}
X.~Jiang, C.~Han, and N.~Mesgarani, ``Dual-path mamba: Short and long-term bidirectional selective structured state space models for speech separation,'' \emph{arXiv preprint arXiv:2403.18257}, 2024.

\bibitem{wang2022self}
S.~Wang, A.~Politis, A.~Mesaros, and T.~Virtanen, ``Self-supervised learning of audio representations from audio-visual data using spatial alignment,'' \emph{IEEE Journal of Selected Topics in Signal Processing}, vol.~16, no.~6, pp. 1467--1479, 2022.

\bibitem{tan2020audio}
K.~Tan, Y.~Xu, S.-X. Zhang, M.~Yu, and D.~Yu, ``Audio-visual speech separation and dereverberation with a two-stage multimodal network,'' \emph{IEEE Journal of Selected Topics in Signal Processing}, vol.~14, no.~3, pp. 542--553, 2020.

\bibitem{michelsanti2021overview}
D.~Michelsanti, Z.-H. Tan, S.-X. Zhang, Y.~Xu, M.~Yu, D.~Yu, and J.~Jensen, ``An overview of deep-learning-based audio-visual speech enhancement and separation,'' \emph{IEEE/ACM Trans. on Audio, Speech and Language Processing}, vol.~29, pp. 1368--1396, 2021.

\bibitem{gao2021visualvoice}
R.~Gao and K.~Grauman, ``Visualvoice: Audio-visual speech separation with cross-modal consistency,'' in \emph{Proc. of Int. Conf. on Computer Vision and Pattern Recognition}.\hskip 1em plus 0.5em minus 0.4em\relax IEEE, 2021, pp. 15\,490--15\,500.

\bibitem{jung2024flowavse}
C.~Jung, S.~Lee, J.-H. Kim, and J.~S. Chung, ``Flowavse: Efficient audio-visual speech enhancement with conditional flow matching,'' \emph{arXiv preprint arXiv:2406.09286}, 2024.

\bibitem{zhangemnlp}
X.~Zhang, D.~Liu, H.~Liu, Q.~Zhang, H.~Meng, L.~P. Garcia~Perera, E.~Chng, and L.~Yao, ``Speaking in wavelet domain: A simple and efficient approach to speed up speech diffusion model,'' in \emph{Proceedings of the 2024 Conference on Empirical Methods in Natural Language Processing}, Y.~Al-Onaizan, M.~Bansal, and Y.-N. Chen, Eds., Nov. 2024, pp. 159--171.

\bibitem{richter2023audio}
J.~Richter, S.~Frintrop, and T.~Gerkmann, ``Audio-visual speech enhancement with score-based generative models,'' in \emph{Speech Communication; 15th ITG Conference}, 2023, pp. 275--279.

\bibitem{mira2023voce}
R.~Mira, B.~Xu, J.~Donley, A.~Kumar, S.~Petridis, V.~K. Ithapu, and M.~Pantic, ``La-voce: Low-snr audio-visual speech enhancement using neural vocoders,'' in \emph{Proc. of {IEEE} Int. Conf. on Audio, Speech and Signal Processing}.\hskip 1em plus 0.5em minus 0.4em\relax IEEE, 2023, pp. 1--5.

\bibitem{zhuvision}
L.~Zhu, B.~Liao, Q.~Zhang, X.~Wang, W.~Liu, and X.~Wang, ``Vision mamba: Efficient visual representation learning with bidirectional state space model,'' in \emph{Proc. of Int. Conf. on Machine Learning}, 2024.

\bibitem{liu2024vmambavisualstatespace}
Y.~Liu, Y.~Tian, Y.~Zhao, H.~Yu, L.~Xie, Y.~Wang, Q.~Ye, and Y.~Liu, ``Vmamba: Visual state space model,'' 2024.

\bibitem{Erdogan2015PhasesensitiveAR}
H.~Erdogan, J.~R. Hershey, S.~Watanabe, and J.~L. Roux, ``Phase-sensitive and recognition-boosted speech separation using deep recurrent neural networks,'' \emph{Proc. of {IEEE} Int. Conf. on Audio, Speech and Signal Processing}, pp. 708--712, 2015.

\bibitem{gong2023contrastive}
Y.~Gong, A.~Rouditchenko, A.~H. Liu, D.~Harwath, L.~Karlinsky, H.~Kuehne, and J.~R. Glass, ``Contrastive audio-visual masked autoencoder,'' in \emph{Proc. ofInt. Conf. on Learning Representations}, 2023.

\bibitem{7178964}
V.~Panayotov, G.~Chen, D.~Povey, and S.~Khudanpur, ``Librispeech: An asr corpus based on public domain audio books,'' in \emph{Proc. of {IEEE} Int. Conf. on Audio, Speech and Signal Processing}, 2015, pp. 5206--5210.

\bibitem{zhao2019sound}
H.~Zhao, C.~Gan, W.-C. Ma, and A.~Torralba, ``The sound of motions,'' in \emph{Proc. of Int. Conf. on Computer Vision}, 2019, pp. 1735--1744.

\bibitem{ephrat2018looking}
A.~Ephrat, I.~Mosseri, O.~Lang, T.~Dekel, K.~Wilson, A.~Hassidim, W.~T. Freeman, and M.~Rubinstein, ``Looking to listen at the cocktail party: A speaker-independent audio-visual model for speech separation,'' \emph{arXiv preprint arXiv:1804.03619}, 2018.

\bibitem{audioset}
J.~F. Gemmeke, D.~P.~W. Ellis, D.~Freedman, A.~Jansen, W.~Lawrence, R.~C. Moore, M.~Plakal, and M.~Ritter, ``Audio set: An ontology and human-labeled dataset for audio events,'' in \emph{Proc. of {IEEE} Int. Conf. on Audio, Speech and Signal Processing}, 2017, pp. 776--780.

\bibitem{mpsenet}
Y.-X. Lu, Y.~Ai, and Z.-H. Ling, ``{MP-SENet: A Speech Enhancement Model with Parallel Denoising of Magnitude and Phase Spectra},'' in \emph{Proc. INTERSPEECH}, 2023, pp. 3834--3838.

\bibitem{chen22vam}
C.~Chen, R.~Gao, P.~Calamia, and K.~Grauman, ``Visual acoustic matching,'' in \emph{Proc. of Int. Conf. on Computer Vision and Pattern Recognition}, 2022.

\bibitem{hou2018audio}
J.-C. Hou, S.-S. Wang, Y.-H. Lai, Y.~Tsao, H.-W. Chang, and H.-M. Wang, ``Audio-visual speech enhancement using multimodal deep convolutional neural networks,'' \emph{IEEE Transactions on Emerging Topics in Computational Intelligence}, vol.~2, no.~2, pp. 117--128, 2018.

\bibitem{10508446}
H.~Chen, Q.~Wang, J.~Du, B.-C. Yin, J.~Pan, and C.-H. Lee, ``Optimizing audio-visual speech enhancement using multi-level distortion measures for audio-visual speech recognition,'' \emph{IEEE/ACM Transactions on Audio, Speech, and Language Processing}, vol.~32, pp. 2508--2521, 2024.

\bibitem{Zheng_2023}
R.-C. Zheng, Y.~Ai, and Z.-H. Ling, ``Incorporating ultrasound tongue images for audio-visual speech enhancement through knowledge distillation,'' in \emph{INTERSPEECH 2023}, p. 844–848.

\bibitem{pesq1}
A.~Takahashi, A.~Kurashima, C.~Morioka, and H.~Yoshino, ``Objective quality assessment of wideband speech by an extension of itu-t recommendation p. 862.'' in \emph{INTERSPEECH}, 2005, pp. 3153--3156.

\bibitem{estoi}
J.~Jensen and C.~H. Taal, ``An algorithm for predicting the intelligibility of speech masked by modulated noise maskers,'' \emph{IEEE/ACM Trans. on Audio, Speech and Language Processing}, vol.~24, no.~11, pp. 2009--2022, 2016.

\bibitem{lin2024VoxBlink}
Y.~Lin, X.~Qin, G.~Zhao, M.~Cheng, N.~Jiang, H.~Wu, and M.~Li, ``Voxblink: A large scale speaker verification dataset on camera,'' in \emph{Proc. of {IEEE} Int. Conf. on Audio, Speech and Signal Processing}, 2024, pp. 10\,271--10\,275.

\end{thebibliography}


\end{document}